\documentclass[preprint]{aastex}
\usepackage{lscape}
\usepackage{ulem}
\usepackage{xcolor}
\newcommand{\caii}{\ion{Ca}{2}}
\newcommand{\fei}{\ion{Fe}{1}}
\newcommand{\feii}{\ion{Fe}{2}}
\newcommand{\hmion}{H$^-$}

\newcommand{\cas}{$Ca$-$s$}
\newcommand{\caw}{$Ca$-$w$}
\newcommand{\str}{Str\"omgren}

\newcommand{\ebv}{$E(B-V)$}

\newcommand{\teff}{$T_{\rm eff}$}
\newcommand{\dteff}{$\Delta T_{\rm eff}$}
\newcommand{\vvhb}{$V-V_{\rm HB}$}
\newcommand{\vhb}{$V_{\rm HB}$}
\newcommand{\vbump}{$V_{\rm bump}$}
\newcommand{\fehi}{[Fe/H]$_{\rm I}$}
\newcommand{\fehii}{[Fe/H]$_{\rm II}$}
\newcommand{\feh}{[Fe/H]}
\newcommand{\dfeh}{$\Delta$[Fe/H]}
\newcommand{\dfehi}{$\Delta$[Fe/H]$_{\rm I}$}
\newcommand{\dfehii}{$\Delta$[Fe/H]$_{\rm II}$}
\newcommand{\tefflogg}{$T_{\rm eff}$ -- $\log g$}
\newcommand{\afei}{$A_{\rm Fe I}$}
\newcommand{\afeii}{$A_{\rm Fe II}$}
\newcommand{\rew}{$\log (W_{\lambda}/\lambda$)}

\begin{document}

\title{ON THE METALLICITY DISTRIBUTION OF THE PECULIAR GLOBULAR CLUSTER M22}

\author{Jae-Woo Lee\altaffilmark{1}}

\altaffiltext{1}{Department of Physics and Astronomy,
Sejong University, 209 Neungdong-ro, Gwangjin-Gu, Seoul, 05006, Korea;
jaewoolee@sejong.ac.kr, jaewoolee@sejong.edu}

\begin{abstract}
In our previous study, we showed that the peculiar globular cluster (GC) M22 
contains two distinct stellar populations, namely the \caw\ and \cas\ groups
with different physical properties, having different chemical compositions, 
spatial distributions and kinematics.
We proposed that M22 is most likely formed via a merger of two GCs
with heterogeneous metallicities
in a dwarf galaxy environment and accreted later to our Galaxy.
In their recent study,  Mucciarelli et al. claimed that M22 is 
a normal mono-metallic globular cluster without any perceptible 
metallicity spread among the two groups of stars, which challenges 
our results and those of others.
We devise new strategies for the local thermodynamic equilibrium (LTE) 
abundance analysis  of red giant branch (RGB) stars in GCs and
show there exists a spread in the iron abundance distribution in M22.

\end{abstract}



\section{INTRODUCTION}
During the last decade, there has been a dramatic paradigm shift 
on the the definition of the GC systems.
Despite the formerly accepted idea of chemical homogeneity,
the variations in the lighter elemental abundances in several GCs in our Galaxy 
had been known for several decades.
The Lick-Texas Group is one of those who undertook the systematic 
study of lighter elemental abundances in several GCs in our Galaxy since 1990's
\citep[e.g.][]{sneden91,kraft94}.
Thanks to the advent of high performance multi-object high-resolution
spectrographs mounted on the large aperture telescopes, 
it is now possible for us to look into the detailed substructure 
of elemental abundance distributions
of the Milky Way GC systems \citep[e.g.][]{carretta09}.
The decades-long lighter elemental variation issue in GC stars
is now considered to be a generic feature of normal GCs
in our Galaxy, most likely engraved during the multi-phase normal GC formation
\citep[e.g.][]{dantona07,decressin07,dercole08}.

Contrary to the normal GC system, one of the key features of the peculiar GCs, 
such as $\omega$ Cen, is the spread or the distinctive substructure 
in the metallicity distributions \citep[e.g.][]{ywlocen,johnson10},
where the heavy elements must have been supplied by supernovae (SNe).
To retain ejecta from energetic SNe explosions,
such peculiar GC systems must have been much more massive in the past 
and they are generally thought to be the remaining 
core of a disrupted dwarf galaxy and accreted to our Galaxy later in time,
expected from the hierarchical merging paradigm in the $\Lambda$CMD 
cosmological model \citep[e.g.][]{sz78,freeman93,moore99}.
The existence of these peculiar GCs have major implications 
in the context of near field cosmology \citep[e.g.][]{bf14};
Are they one of the original building blocks to our Galaxy,
mitigating so-called  the ``missing satellite problem?'' \citep{moore99};
How did they relate to the formation of the Galactic halo and 
numerous streamers? \citep[e.g.][]{belokurov06}.
These are examples of the outstanding problems 
that we have to challenge in the next decade.

To measure metallicity of stars in GCs from high-resolution spectroscopy, 
the LTE (local thermodynamic equilibrium) analysis is being widely used 
for the sake of convenience, where the final results critically depend on 
the stellar atmosphere models with a few appropriate input stellar parameters, 
such as effective temperature, surface gravity and turbulent velocity, 
and the oscillator strengths of the absorption lines.
Although simple, the derivation of stellar elemental abundances is not a
trivial task even for nearby bright stars. The recent study of 
\citet{baines10} may highlight the current situation. They showed that
the interferometric effective temperatures for nearby K giant stars
do not agree with those from spectroscopic observations, suggesting
a missing source of opacities in stellar atmosphere models. The
situation would be even worse for fainter stars, such as those in
GCs.
Another line of difficulty is that changes in surface gravity can mimic
the chemical compositions in the regime of RGB stars in GCs
\citep[for example, see][]{gray08}, 
where \hmion\ is the major source of the continuum opacity 
and the \hmion\ population varies with the electron pressure, therefore,
the surface gravity of RGB stars.

The spread in the metallicity distribution of M22 has been 
a controversial topic for many years. The recent several studies of the cluster
have found that M22 has a bimodal heavy elemental abundance distribution
\citep[e.g.,][]{dacosta09,jwlnat,lee15,marino09,marino11,marino12,marino13}.
The high-resolution spectroscopic elemental abundance measurements 
of RGB stars in the peculiar GC M22
by \citet{brown92}\footnote{See also Figure~3 of \citet{jwlnat},
where we showed a bimodal heavy elemental abundance distribution including 
iron of M22 RGB stars using the results of \citet{brown92}.}
and \citet{marino09,marino11} showed a distinctive bimodal metallicity 
distribution.
Their results were based on the spectroscopic \teff\ and $\log g$, 
which require the excitation and ionization equilibria,
\begin{eqnarray}
 \frac{\partial A_{\rm Fe I}}{\partial \chi_{\rm exc}} &=& 0, \nonumber
\end{eqnarray}
and
\begin{eqnarray}
A_{\rm Fe I} &=& A_{\rm Fe II}, \nonumber
\end{eqnarray}
where \afei\ and  \afeii\ are iron abundances from
\ion{Fe}{1} and \ion{Fe}{2} lines and $\chi_{\rm exc}$ 
is the excitation potential.

It has been frequently suspected that the iron abundance from 
the \ion{Fe}{1} line suffers from non LTE (NLTE) effects \citep[see, 
for example,][]{ti99,ki03,m68,n5694,lee10}. 
Since metal-poor stars have much weaker metal absorption in the ultraviolet,
more non-local ultraviolet flux can penetrate from the deeper layers.
This flux is vital in determining the ionization equilibrium of the atoms,
resulting in deviations from LTE \citep{ti99,ki03,m68}.
In this regard, the traditional spectroscopic surface gravity
determination method may be in error, which led \citet{ki03}
to propose to use the photometric gravity for elemental abundance
study of RGB stars in GCs, where the bolometric correction
could be the dominant source of uncertainty for the photometric gravity.
In principle, the use of the \fehii\ abundance as for the metallicity
scale of RGB stars in GCs is most likely an appropriate approach
since \feii\ is by far the dominant species
and, therefore, the number of \feii\ atoms is unaffected by the NLTE effect.
In practice, however, the \fehii\ abundances of RGB stars sensitively depends
not only on the surface gravity and effective temperature but also on
the metallicity of the input atmosphere model, 
which also affects the continuum opacity.
We will show later that an iterative procedure is useful to reduce
the error raised by the incorrect metallicity of the input atmosphere model.

In their recent study, \citet[][Mu15 hereafter]{mucciarelli15} re-analyzed 
M22 RGB stars of \citet{marino11} using three different approaches: 
(1) Spectroscopic \teff\ and $\log g$ (Method 1); 
(2) spectroscopic \teff\ and photometric $\log g$ (Method 2);
and (3) photometric \teff\ and $\log g$ (Method 3).
They confirmed a bimodal iron abundance distribution of M22 RGB stars
by \citet{marino09,marino11},
when they relied on spectroscopic \teff\ and $\log g$ (Method 1).
Oddly enough, when they used photometric $\log g$ 
(Method 2 and 3 by Mu15),
the allegedly well-established bimodal iron distribution of M22
disappeared in \fehii.
Using the photometric gravity is most likely a correct approach but
how can this be interpreted?
We will show later that it is likely that Mu15
used incorrect surface gravity and the metallicity of input atmosphere models
in their analysis and the separation in \fehii\ can be brought out
more fully if different methods to compute these parameters are used.

In this paper, we revisit the internal metallicity distribution of M22.
We developed new methods to estimate the surface gravity and we found that
there exists substantial metallicity difference
between the \caw\ and \cas\ groups\footnote{The \caw\ (calcium-weak) and 
the \cas\ (calcium-strong) groups are defined to be RGB stars with smaller 
or larger $hk$ index values, respectively, at a given $V$ magnitude.
They are equivalent to the $s$-process-poor (\caw) and 
the $s$-process-rich (\cas) groups classified by \citet{marino11}.}
in M22 \citep{jwlnat,lee15}. 
Throughout this paper, metallicity refers to \fehii, unless specified.

\section{LESSONS LEARNED FROM PREVIOUS STUDIES}\label{sec2}
Before turning to the metallicity distribution of M22, we would like to
revisit the critical issues on the LTE analysis; surface gravity, 
effective temperature and the metallicity of the input atmosphere model.

\subsection{\fehii\ with surface gravities independent of ionization equilibrium}
As mentioned above, \citet{ki03} suggested to use \fehii\ derived from
photometric gravity as metallicity of RGB stars.
They pointed out that \fehii\ is essentially independent of NLTE effects,
such as \ion{Fe}{1} overionization by non-local UV flux since 
\ion{Fe}{2} is the dominant species in GC RGB stars.

\begin{figure}
\epsscale{1}
\figurenum{1}
\plotone{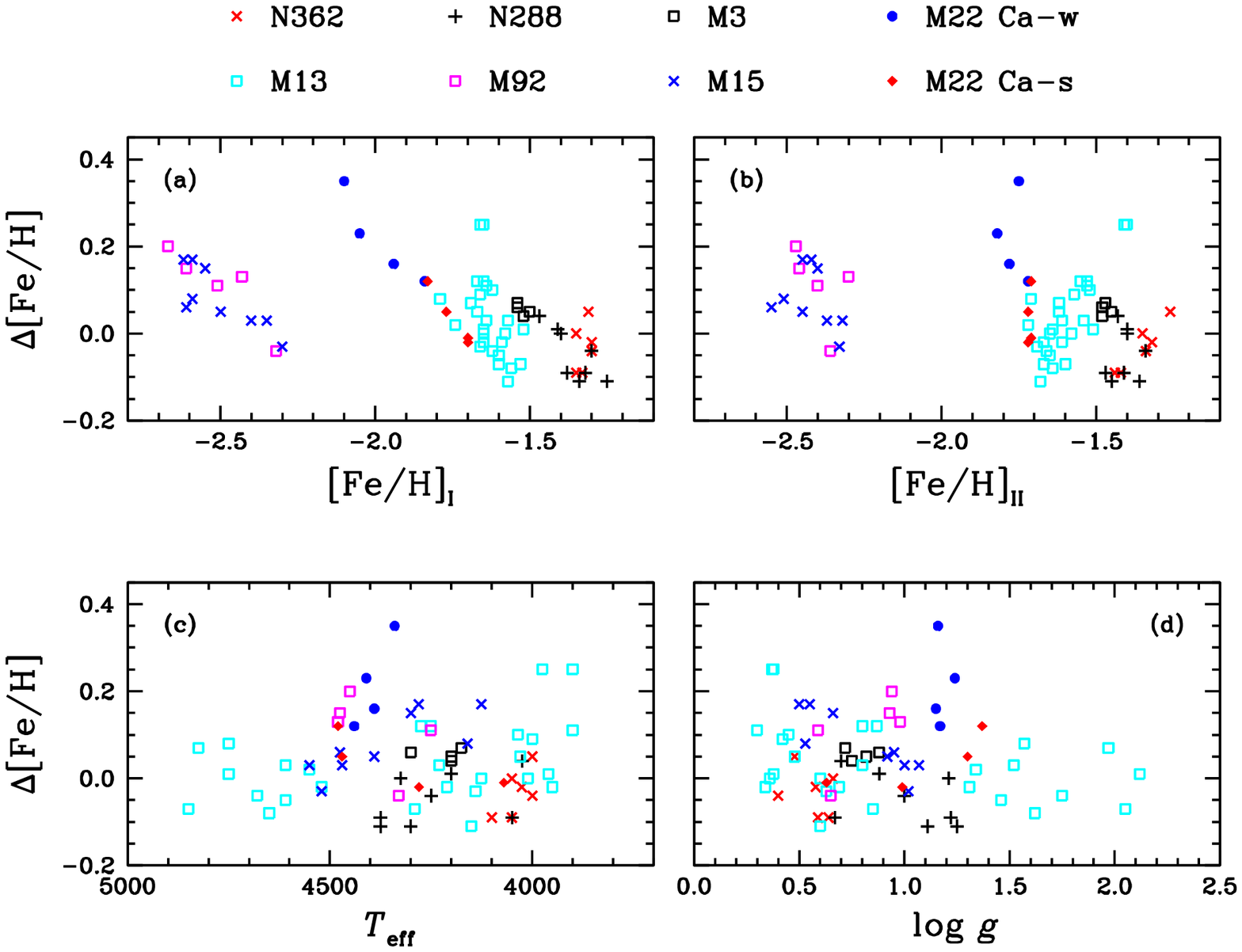}
\caption{Differences in \fehi\ and \fehii\ 
from photometric gravities of RGB stars in six GCs studied by \citet{ki03}.
We also show \dfeh\ of four \caw\ and four \cas\ M22 RGB stars
using \fehi\ and \fehii\ measurements from Method 2 of Mu15. 
Note that M22 \caw\ stars have preferentially
larger \dfeh\ values than RGB stars in other clusters do.
}\label{fig:ki03}
\end{figure}

In Figure~\ref{fig:ki03}, we show plots of \dfeh\ against \fehi, \fehii,
\teff\ and $\log g$ for the six Group 1 clusters\footnote{Clusters with 
\teff\ and $\log g$ based on colors and absolute magnitudes.} of \citet{ki03}.
Of particular concern is Figure~\ref{fig:ki03} (a) and (b), where each GC is showing 
its own correlation between \fehi\ versus \dfeh\ (or \fehii\ versus \dfeh).

In Figure~\ref{fig:fit}, we show plots of $\log g$ versus \teff, 
\teff\ versus \vvhb, and $\log g$ versus \vvhb\ for the six Group 1 GCs.
Also shown are Victoria-Regina model isochrones for 12 Gyr \citep{vr}.
To estimate the \vhb\ level of the model isochrones, 
we used our previous relation
\citep{n6723},
\begin{equation}\label{eq:vhb}
M_V\mathrm{(RR)} = (0.214 \pm 0.047)(\mathrm{[Fe/H]} + 1.5) + (0.52 \pm 0.13).
\end{equation}
Note that this metallicity-luminosity relation for RR Lyrae variables gives 
$(m-M)_0$ = 18.54 $\pm$ 0.13 mag for LMC.
It also should be emphasized that the adopted age of the model isochrones
does not affect our results presented in this work.

In plots of $\log g$ versus \teff\ and \teff\ versus \vvhb, the loci of
the model isochrones are in excellent agreement with the observations.
It should not be a surprise that, at a given temperature, the surface
gravity of the metal-poor stars is lower and \vvhb\ is smaller 
than those of the metal-rich stars.
This implies that applying a single \teff\ versus $\log g$ relation
for groups of stars with heterogeneous metallicity may result in incorrect 
surface gravity estimates and, as a consequence, incorrect elemental abundances. 
This is especially true when one adopt \fehii\ as metallicity
of GC RGB stars. The \feii\ line opacity does not vary with surface
gravity since the almost all iron atoms are populated in the first ionized level.
On the other hand, the \hmion\ continuum opacity is sensitively dependent
on the electron pressure and, therefore, surface gravity. 
As a consequence, in the regime of the  linear part of the curve of growth, 
at fixed metallicity, the equivalent widths (EWs) of \feii\ lines 
increases with decreasing surface gravity.
In the figure, the surface gravity dependency on the metallicity of 
the metal-poor RGB stars at a fixed temperature is given by
\begin{equation}\label{eq:fehlogg}
 \frac{\partial {\rm [Fe/H]}}{\partial \log g} \approx 1.8\ {\rm dex}.
\end{equation}
In other words, a change in the surface gravity by 0.1 in the cgs unit
corresponds to a change in metallicity by about 0.2 dex in the figure.

On the other hand, a rather tight relation in $\log g$ versus \vvhb\ 
can be found, suggesting that the \vvhb\ magnitude can be a reddening- 
and distance-independent  surface gravity indicator for RGB stars in GCs.
\citep[see also,][for the use of $K - K_{\rm HB}$ as a temperature indicator
of heavily reddened metal-rich GC Palomar~6]{pal6}.
But caution is advised that the observed \vvhb\ magnitude can be vulnerable to
the foreground differential reddening effect.

\subsection{The utility of spectroscopic effective temperature}
As mentioned previously, \citet{ki03} suggested to use the photometric
surface gravity. 
However, ironically, under the condition that LTE is not valid,
they concluded to use spectroscopic temperature from \fei\ lines.
We would like to discuss the utility of the spectroscopic 
effective temperature under the assumption that the conclusion made
by \citet{ki03} is valid.

In Figure~\ref{fig:log}, we show plots of the \afei\ against 
the excitation potential for NGC~6752-mg10 by \citet{yong13}.
As they noted, the quality of their spectra is superb; 
a resolving power of $R$ = 110,000 and S/N $\geq$ 150 per pixel
near 5140 \AA.
To perform a LTE abundance analysis, we used the 2014 version of 
the stellar line analysis program MOOG \citep{moog}
and we interpolated $\alpha$-enhanced Kurucz atmosphere models
with new opacity distribution functions using a FORTRAN program 
kindly provided by Dr.\ McWilliam (2005, private communication).
Using the weak \ion{Fe}{1} lines, \rew\ $\leq  -5.2$, 
we derived a spectroscopic temperature for this star 
by forcing the condition of the excitation equilibrium,
i.e.\ $\partial$\afei/$\partial\chi$ = 0, and we obtained 
the effective temperature of 4275 K.
Note that our effective temperature for this star is slightly cooler than 
those from the line-by-line differential analysis  
with respect to NGC~6752-mg9 and NGC~6752-mg6
by \citet{yong13}, 4291 K and 4295 K, respectively.
We derived the linear fits to the data, assuming 
\afei\ $\propto$  slope$\times\chi$,
and we show them with thin dashed lines in the figure.
In the parentheses in each panel, we also show the $\log g$ value 
in the cgs unit, the slope in the excitation potential versus 
the iron abundance, \fehi\ and \fehii. 
As can be seen, once the spectroscopic effective temperature 
is correctly determined,
the assumption of the excitation equilibrium holds for rather wide range 
of the surface gravity, $\Delta\log g \approx$ 1.20 in this case.
Therefore, it can be concluded that the slightly incorrect input 
surface gravity does not significantly affect the spectroscopic effective 
temperature in our results presented here.

\subsection{Iterative derivation of metallicity: 
\fehi\ and \fehii\ versus input atmosphere model parameters}
It is well-known fact that the inferred metallicity of GC RGB stars from
high resolution spectroscopy
is critically dependent on the stellar parameters of the input atmosphere model.
Here, we show how \fehi\ and \fehii\ behave against the changes in the input 
stellar parameters. We also would like to demonstrate the importance of 
the iterative derivation of the metallicity.

In Figure~\ref{fig:fehvspam}, we show \fehi\ and \fehii\ of 9 RGB stars in NGC~6752 
by \citet{yong13}. In each panel, the red crosses denote our spectroscopic 
\fehi\ and \fehii\, which were derived using the weak \fei\ and \feii\ lines
measured by \citet{yong13}. 
It should be noted that our \feh\ values are consistent with those of
\citet{yong13}. 
Using our spectroscopic \teff, $\log g$, and \feh\ as reference grids,
we examine how the changes in the stellar parameters of the input
atmosphere models affect the resultant metallicity.
We run MOOG using the Kurucz atmosphere models, whose stellar parameters 
are different from the reference grids by $\Delta$\teff\ = $\pm$200 K,
$\Delta$\feh\ = $\pm$1.0, and $\Delta\log g$ = $\pm$0.3.
We use the \fehii\ abundances returned from MOOG as the reference metallicity
and we calculate new atmosphere models, with which we run MOOG again.
We iterate this process five times and we show our results 
in Figure~\ref{fig:fehvspam} with blue solid lines.

The figure shows that the inferred \fehi\ abundances with sufficient numbers
of iterations are only affected by the changes in the effective temperature.
As shown, \fehi\ abundances of individual stars against the changes in
the metallicity and the surface gravity of the input model atmosphere
converge to their spectroscopic \fehi\ values within 2 or 3 iterations.
However, the iterations with the effective temperature offsets of $\pm$200K
fail to regain spectroscopic \fehi,
implying incorrect \teff\ estimate results in irrecoverable deviations
in the derived metallicity.

The \fehii\ behaves differently against the changes in input stellar parameters.
The inferred \fehii\ abundances with the effective temperature offsets shifted 
in the opposite direction of the changes in the \fehi\ abundance.
A rather simple explanation of the temperature effect in cool stars
can be found in \citet{gray08}, for example.
Similar to \fehi, the spectroscopic metallicity can be regained
with the iterative derivation of the \fehii\ abundances against the changes
in the metallicity of the input atmosphere model but, however,
the difference in \fehii\ between the spectroscopic metallicity and
that from the first iteration is much larger than that can be seen in \fehi. 
As we have already discussed earlier, the \fehii\ abundance sensitively depends
on the surface gravity of the input atmosphere model. 
The Figure shows that the spectroscopic metallicity can not be regained
with incorrect surface gravity.

We also performed the same procedures using the \fehi\ abundances of individual
stars as reference metallicity to calculate the input atmosphere model
used in the next iteration and we obtained the same results shown above.

Figure~\ref{fig:delfeh} summarize our exercise.
For both \fehi\ and \fehii, the inferred metallicity from the first iteration
with offsets in the stellar parameters could be very different from
those with correct stellar parameters. 
However, some discrepancies vanish after 2 or 3 iteration processes.

Our exercise demonstrates that, with the iterative procedure, the \fehi\ abundance
depends only on the effective temperature, while the \fehii\ abundance depends
both on the effective temperature and the surface gravity.
It also shows the importance of having correct stellar parameters,
especially for the LTE analysis of \fehii\ abundances, 
as we have discussed earlier.

\section{PREVIOUS EVIDENCE OF THE BIMODAL METALLICITY DISTRIBUTION OF M22}\label{sec3}
The vivid evidence of the bimodal metallicity distribution of RGB stars in M22 
from narrow-band photometry and high-resolution spectroscopy can be found
in \citet{jwlnat}, \citet{lee15} and \citet{marino09,marino11}. 
In our earlier study of the cluster, we extensively discussed that
there are several observational lines of evidence that cannot be easily 
explained without invoking a bimodal metallicity distribution between 
two groups of stars, namely the \caw\ and the \cas\ groups
as shown in Figure~\ref{fig:m22cmd}
\citep[e.g., see Figures~9, 13 -- 21 of][and references therein]{lee15};

\begin{itemize}
\item The \caii\ H\&K absorption strengths RGB stars 
at a given $V$ magnitude in M22
from both narrow-band photometry \citep{jwlnat,lee15} and
low-resolution spectroscopy \citep{norris83,lim15}
show a bimodal distribution.

\item The infrared \caii\ triplet by \citet{dacosta09} 
also show a bimodal distribution among RGB stars in M22.

\item The $m1$ versus $V$ CMD as shown in Figure~\ref{fig:m22cmd}
\citep[see also Figure~19 of][]{marino11}
also requires a bimodal metallicity distribution in M22 RGB stars. 
The variation in the lighter elements only, such as CNO, cannot explain 
this distinct double $m1$ RGB sequences of M22.
The differential foreground reddening effect cannot reproduce
the observed multi-color CMDs accordingly \citep{lee15}.

\item  The $V$ magnitude of the RGB bump, \vbump, of the \cas\ group is 
significantly fainter than that of the \caw\ group, which
strongly suggests that the \cas\ group is more metal-rich than
the \caw\ group is.
The difference in the \vbump\ between the two groups cannot be explained 
by the differential foreground reddening effect \citep{lee15}.

\item  The slope of the \cas\ RGB stars in the $cy$ versus $V$ CMD is
significantly larger than that of the \caw\ RGB stars, indicative
of the metal-rich nature of the \cas\ group \citep{lee15}.

\item  The CN-CH positive correlation superposed on two separate CN-CH
anticorrelations \citep{lim15} can be expected naturally if M22 is composed of 
two groups of stars with heterogeneous metallicities \citep{lee15}.

\item Finally, another piece of evidence can be found in the metallicity 
distribution of the blue horizontal branch (BHB) stars in M22 \citep{marino13}.
In Figure~\ref{fig:hb}, we show the metallicity distributions
for M22 BHB stars.
Both the LTE and the NLTE treatments show
the similar degree of metallicity spreads in \fehi\ and \fehii\
(six stars for \fehi\ and seven stars for \fehii).
As shown in Table~3 of \citet{marino13}, their derived \fehii\ value
is less sensitive to the effective temperature and to the surface gravity, 
$\Delta$\fehii = $\pm$0.06 dex and $\pm$0.01 dex
for $\Delta$\teff\ = $\pm$170 K and $\Delta \log g$ = $\pm$0.20 respectively,
and we are likely seeing the real metallicity spread in M22 BHB stars.

\end{itemize}

\section{NO METALLICITY SPREAD IN M22?}\label{sec4}
As mentioned above, Mu15 re-analyzed 
M22 RGB stars of \citet{marino11} using three different approaches. 
We show 17 RGB and asymptotic giant branch (AGB) stars studied by Mu15 
in Figure~\ref{fig:m22cmd}.
Among 12 RGB stars tagged by Mu15, six RGB stars belong
to each of two RGB groups, according to our previous classification of
RGB stars based on the $hk$ index at a given $V$ magnitude. 

Figure~\ref{fig:pam} shows a plot of the $\log$ \teff\ versus $\log g$ of 
spectroscopic target stars, where the filled symbols are 
for Method 1 (spectroscopic \teff\ and $\log g$) and the open symbols 
are for Method 2 (spectroscopic \teff\ and 
photometric $\log g$) of Mu15.
Also shown are Victoria-Regina model isochrones for 12 Gyr with 
[Fe/H] = $-$1.40, $-$1.70 and  $-$2.00 dex \citep{vr}.
The most conspicuous feature of the figure is that  the positions of 
stars from Method 2 by  Mu15 are 
rather well aligned in a narrow strip between model isochrones 
with [Fe/H] = $-$1.40 and $-$1.70 dex on the \tefflogg\ plane, 
which is  most likely due to the adopted {\em single} 
\tefflogg\ relation by Mu15.
It should be worth to point out that, because the CMD of Mu15 has
a single RGB sequence, their photometric effective temperature
and surface gravity will also define a single isochrone.
The broad band colors adopted by them are not sensitive
to small \feh\ differences.
On the other hand, positions of stars from Method 1 occupy rather wide ranges
in surface gravity at given effective temperature.

We performed a LTE abundance analysis using EWs of M22 RGB stars
measured by Mu15 to calculate the the mean iron abundance dependence
on model atmospheres. In Table~\ref{tab:dep}, we show our result 
for eight RGB stars in M22 with 0.5 $\leq \log g_{\rm phot} \leq$ 1.5,
where $\log g_{\rm phot}$ is the photometric surface gravity in the cgs unit.
As shown in the table, the change in the surface gravity by
$\delta\log g \approx$ 0.2 results in $\Delta$\fehii\ $\approx$ 0.1 dex
at the fixed \teff, in the sense that the \fehii\ value increases 
with the surface gravity.\footnote{Note that the metallicity dependency
on the surface gravity from the LTE analysis is about four times smaller 
than that from the isochrones as in Equation~\ref{eq:fehlogg}.}
If we take this at face value, 
a single \tefflogg\ relation, which is valid only for mono-metallic
stellar systems such as normal GCs in our Galaxy, may be responsible for 
the narrow uni-modal \fehii\ distribution of M22 RGB stars as claimed 
by Mu15 as shown in their Figure~2. 

In Figure~\ref{fig:method2}, we show the metallicity distribution
of M22 RGB stars using Table~4 of Mu15 (Method 2).
Note again that we use 8 RGB stars with 0.5 $\leq \log g_{\rm phot} \leq$ 1.5 only
(corresponding to 11.5 $\leq V \leq$ 12.75 mag in Figure~\ref{fig:m22cmd})
to avoid the potential effect raised by very different surface gravity 
in the stellar atmosphere model calculations.
According to our population classification scheme for M22 based on
our $hk$ index at a given $V$ magnitude, stars 61, 71, 200068 and 200076 
belong to the \caw\ group, while stars 51, 88, 200025 and 200101 
belong to the \cas\ group.
Figure~\ref{fig:method2} (a) shows that each RGB group has different
mean iron abundance both in \fehi\ and \fehii. 
It is very interesting to note that the differences in the mean iron abundances 
between the two groups are larger than a 2.5 $\sigma$ level both 
in \fehi\ and \fehii,
although the separation in the mean \fehii\ values is as small as 
$\approx$ 0.05 dex.
In Table~\ref{tab:ks}, we show our results (Mu15 M2).

Figure~\ref{fig:method2} (b) -- (e) show the cumulative metallicity
distributions and the generalized histograms of the metallicity 
distributions for each group.
As shown, the \fehi\ distribution show two distinct peaks, 
while that for \fehii\ shows a broad single peak
in the overall metallicity distribution,
similar to those obtained by Mu15.
But it should be emphasized that the \fehii\ distribution by 
Mu15 is composed of two separate mono-metallic distributions; 
only the mean \fehii\ values from the Method 2 by Mu15
for each group of stars happen to be similar.
We performed a Student's $t$-test to see if the metallicity 
distributions of the two groups of stars are identical.
We obtained that the significance levels to reject the hypothesis that
the mean \feh\ values of the \caw\ and the \cas\ groups are identical,
are 1.93 \% and 8.98 \% for \fehi\ and \fehii, respectively.
We also performed a randomization test. The significance levels 
to reject the hypothesis for being an identical metallicity distribution
from bootstrap method are 0.00 \% for both \fehi\ and \fehii,
strongly suggest that the metallicity distributions for the \caw\ and
the \cas\ groups by Mu15 are not statistically identical.
It should not be a surprise because \citet{lee15} 
already discussed many aspects of heterogeneous nature between 
the two groups as summarized in \S{\ref{sec3}}.

Figure~\ref{fig:method2} (f -- g) shows \dfeh\ (= \fehii\ $-$ \fehi)
against \fehi\ and \fehii\ and they are very intriguing.
It should be reminded that we chose stars with a narrow range of 
$V$ magnitude for both groups to avoid the potential effect raised
by the very different surface gravity.
As shown in the figure, the discrepancies between \fehi\ and \fehii\ of 
the \caw\ group RGB stars are preferentially much greater than 
those of the \cas\ group, reaching as large as \dfeh\ $\approx$ 0.4 dex.
For comparison, we show \dfeh\ of M22 along with six group 1 GCs
by \citet{ki03} in Figure~\ref{fig:ki03}, where the much greater \dfeh\ 
values of the \caw\ RGB stars in M22 can be clearly seen.
If the two groups of stars have the same metallicity and the same surface gravity
so that they suffer similar degree of NLTE effect, one would expect to see the similar 
degree of \dfeh\ for both groups, in sharp contrast to the results of Method 2 by Mu15.
We will show later that, with a mock peculiar GC, the incorrect surface 
gravity estimate by Mu15 and the incorrect metallicity of 
input model atmospheres are responsible for the discrepancy in \dfeh.

Another aspect needs to consider is the initial metallicity used 
during the atmosphere model calculations.
In Figure~\ref{fig:Mu15iter}, we show plots for eight M22 RGB stars, 
similar to Figure~\ref{fig:fehvspam}
for NGC~6752. 
In parentheses of each panel, we show the method for the metallicity derivation
by Mu15 (M1 or M2) and the reference metallicity for the iterations
(I for \fehi\ and II for \fehii).
In the figure, the crosses denote the Mu15's \fehi\ and \fehii\ values
from M1 and M2 methods and blue and red colors denote the \caw\ and \cas\
RGB stars, respectively.
Using \fehi\ and \fehii\ abundances as trial metallicities and
\teff\ and $\log g$ from M1 and M2 methods by Mu15, we performed
the iterative derivations of \fehi\ and \fehii\ for individual stars.
As shown, \fehi\ values do not vary significantly with the number of iterations
with respect to the spectroscopic \feh,
meaning that the effective temperature adopted by Mu15 is correct.
On the other hand, the discrepancy in \fehii\ for \cas\ RGB stars from Method 2
are preferentially larger, indicating that the surface gravities 
of the \cas\ RGB stars adopted by Mu15 is most likely underestimated.
The figure also indicates that the separation in the \fehii\ abundances
between the \caw\ and the \cas\ groups becomes larger with 
the iteration processes.

Finally, comparisons of EWs between the two groups may also help to
elucidate the underlying metallicity distributions of the cluster.
In Figure~\ref{fig:ew}(a), we show the line-by-line EW difference
between stars 88 (\cas) and 200076 (\caw). 
Both stars have similar visual magnitudes and colors,
($V$, $b-y$) = (12.54, 0.89) for the star 88 and (12,39, 0.90)
for the star 200076. 
Therefore, if there is no significant foreground differential reddening
and if the both stars have the same metallicity, they should have
similar \teff, $\log g$ and furthermore the similar EW strengths.
As shown in the figure, the EWs of the \cas\ group star 88 are systematically
larger than those of the \caw\ group stars 200076.
A comparison between the mean EWs of the four \cas\ group stars 
(51, 88, 200025 and 200101) and the four \caw\ group stars 
(61, 71, 200068 and 200076) show the same trend that 
the mean EWs of the \cas\ group are larger than
those of the \caw\ group, 12.0 $\pm$ 0.4 m\AA\ for \ion{Fe}{1} lines
and 1.8 $\pm$ 0.4 m\AA\ for \ion{Fe}{2} lines.
Note that the eight stars above have similar visual magnitudes and colors
and they should have similar \teff, $\log g$ and \feh\ if they belong
to a single stellar population.
A simple explanation why \feii\ lines are less sensitive to changes in metallicity
as follows.
Since both the fraction of \fei\ atoms and the \hmion\ continuum opacity 
of RGB stars depend on the electron pressure (i.e., metallicity or 
surface gravity) and, furthermore, the two effects are expected to cancel out,
the EWs of \fei\ lines grow with metallicity 
at fixed effective temperature.
On the other hand, the \feii\ atoms are the dominant species 
and only the electron pressure has an effect on the \hmion\ continuum opacity,
which has an opposite effect on the growth of EWs with metallicity.
Therefore, the EWs of \feii\ lines grows at a slower rate.

We suspect that the metallicity measurements by Mu15
may be slightly incorrect and we devise new methods to derive metallicity
of RGB stars in appropriate and consistent manners.

\section{M55 + NGC~6752: A MOCK PECULIAR GC}\label{sec5}
In our previous study, \citet{lee15} showed that a combination of two normal
GCs, M55 and NGC~6752, can reproduce many aspects of peculiar photometric
characteristics of M22.
In Figure~\ref{fig:m55n6752cmd}, we show a composite CMDs for M55 and NGC~6752,
which may highlight the importance of the choice of photometric passbands
to distinguish multiple stellar populations in GCs.
For NGC~6752 stars, we add offsets of 0.030, $-$0.010, $-$0.005,
and 0.700 mag in ($b-y$), $m1$, $hk$,\footnote{We adopt the interstellar
reddening law by \citet{att95}; $E(b-y) = 0.74E(B-V)$, $E(m1) = -0.33E(b-y)$,
and $E(hk) = -0.155E(b-y)$.}
and $V$, respectively, 
in order to place NGC~6752 stars in the M55 colors and magnitude scale.
In the figure, we also show RGB stars studied from high-resolution
spectroscopy of the clusters \citep{carretta09uves,yong13}\footnote{
It should be mentioned that the spectral resolving power for the data from
\citet{yong13} is much higher than that of \citet{carretta09uves}, 
110,000 versus 40,000, and the scatter in the mean metallicity for 
NGC~6752 by \citet{yong13} is much smaller than that by \citet{carretta09uves}.
Note that M55-7000020 \citep{carretta09uves} and 
NGC~6752-mg9 \citep{yong13} are most likely AGB stars of the clusters, 
and we do not make use of them in the following analysis.} 
and the Victoria-Regina model isochrones for 12 Gyr with \feh\ = $-$1.84 and $-$1.53.

Despite the difference in metallicity, the RGB sequence of NGC~6752 is 
in excellent agreement with that of M55 in $(b-y)$ versus $V$ CMDs,
and it is difficult to discern different populations.
In such a case, one can be easily misled to adopt 
a single \tefflogg\ relation for heterogeneous stellar populations.  
On the other hand, the distinct double RGB sequences can be clearly
seen in $m1$ versus $V$ and $hk$ versus $V$ CMDs, where the necessity
for double \tefflogg\ relations is obvious.

\subsection{Photometric method using two separate relations}\label{sec5.1}
First, we derived the metallicity distributions of M55 and NGC~6752
by employing two separate \tefflogg\ relations, following the procedure
recommended by \citet{ki03}.
With the EW measurements by \citet{carretta09uves}
for \ion{Fe}{1} and \ion{Fe}{2} lines,
we performed a LTE abundance analysis.
Applying the color-temperature relation and the equation for the bolometric 
correction given by \citet{alonso99}, we calculated the photometric effective 
temperature and the surface gravity using our own \str\ photometry.
During our calculations, we adopted \feh\ = $-$1.95 and $-$1.55 
as the input metallicities for M55 and NGC~6752, 
respectively \citep{carretta09uves}, and we used the distance moduli and 
foreground reddening values by \citet{harris96}.
During our analysis, we used weak lines only, \rew\ $\leq  -5.2$, 
for both \ion{Fe}{1} and \ion{Fe}{2} in order to minimize the effect 
of the adopted micro-turbulent velocity on the derived metallicity.
We show our results in Table~\ref{tab:mock} and Figure~\ref{fig:photorelation}.
Our \fehii\ measurements for the clusters are in good agreement
with those by \citet{carretta09uves}, \dfehii\ = $-$0.01 $\pm$ 0.03 dex for M55
and \dfehii\ = 0.05 $\pm$ 0.02 dex for NGC~6752 in the sense of current work
minus those of \citet{carretta09uves}.
These small differences are thought to be 
mainly due to slightly different final color-temperature relations.
\citet{carretta09uves} used the $(V-K)$ color-temperature relations
given by \citet{alonso99} to derive the initial \teff\ but
they applied their own \teff\ versus $V$ magnitude relation
to derive their final adopted \teff.

In Figure~\ref{fig:photorelation} (b) and (c),
we show \dfeh\ (= \fehii\ $-$ \fehi) against \fehi\ and \fehii, 
where the extent of \dfeh\ for both clusters
are in good agreement with those of six GCs studied by \citet{ki03} 
as shown in Figure~\ref{fig:ki03}.

Figure~\ref{fig:photorelation} (d) shows a plot of \teff\ versus $\log g$ 
along with the Victoria-Regina model isochrones for 12 Gyr with \feh\ = $-$1.84
and $-$1.55 \citep{vr}.
At a given \teff, the stars in the metal-poor cluster M55 have lower surface
gravity, although the mean difference in $\log g$ between M55 and NGC~6752
is not as large as that can be inferred from model isochrones.
Using separate relations, the differences in the mean \fehi\ and \fehii\
are 0.44 $\pm$ 0.02 and 0.47 $\pm$ 0.03, respectively, and they are in excellent
agreement.

Following the same procedure described earlier,
we performed iterative derivations of \fehi\ and \fehii\ for the clusters.
As shown in Figure~\ref{fig:photorelation} and Table~\ref{tab:mock},
\fehi\ and \fehii\ from iterative processes are in good agreement
with those without the iterative process.

We also make use of the EW measurements for the \fei\ and \feii\ lines for
NGC~6752 RGB stars by \citet{yong13}.
We show our results in Figure~\ref{fig:m55n6752comp2} and Table~\ref{tab:mock}.
Note that the oscillator strengths for the individual lines are slightly 
different between those adopted by \citet{carretta09uves} and \citet{yong13}.
For our results presented in Figure~\ref{fig:m55n6752comp2},
we used the oscillator strengths by \citet{yong13} because they provided 
more lines. 
As summarized in Table~\ref{tab:mock}, \fehi\ and \fehii\ abundances using
$gf$-values by \citet{carretta09uves} and \citet{yong13} are slightly different,
but the differences in the mean values are no larger than 0.04 dex.
Therefore, it can be said that the choice of the set of oscillator strengths 
does not affect our primary results presented here.
The differences in \fehi\ 
(\dfehi\ = \fehi$_{\rm , NGC~6752}$ $-$ \fehi$_{\rm , M55}$) 
and \fehii\
(\dfehii\ = \fehii$_{\rm , NGC~6752}$ $-$ \fehii$_{\rm , M55}$) 
are 0.38 - 0.45 dex and 0.44 - 0.50 dex, respectively.

\subsection{Photometric method using a single relation}
Assuming that our mock GC (i.e.\ M55 + NGC~6752) is a mono-metallic GC with 
\feh\ $\approx -1.55$ dex (i.e.\ that of NGC~6752),
we calculated photometric effective temperatures and surface gravities of 
individual stars using the $(b-y)$ color-temperature relation
and the equation for the bolometric correction given by \citet{alonso99}.
Using the weak \ion{Fe}{1} and \ion{Fe}{2} lines only, \rew\ $\leq -5.2$,  
we derived the metallicity of individual stars in both clusters and 
we show our results in Table~\ref{tab:mock} and Figure~\ref{fig:photorelation}.
The mean \fehi\ abundance of M55 remains unchanged, while
that of \fehii\ increases almost 0.15 dex compared to the results
from correct input stellar parameters for M55 presented in \S{\ref{sec5.1}}.
The difference in the effective temperature between the two methods
(i.e.\ two separate relations versus a single relation for M55 and NGC~6752)
is negligibly small, \dteff\ = 12 K, in the sense that the mean effective
temperature of M55 RGB stars is slightly warmer when the correct $(b-y)$ 
color-temperature relation is used. 
As we have discussed earlier, 
the \fehi\ abundance is relatively insensitive to changes
in the surface gravity and the metallicity of the input model atmosphere
since the effects due to change in the number of \fei\ species and
that in \hmion\ continuum opacity are expected to cancel.
However, the surface gravity becomes larger, 
$\Delta \log g$ $\approx$ 0.1, and the metallicity of the input
model atmospheres is higher (from $-$1.95 to $-$1.55 dex)
when a single relation with respect to NGC~6752 is used.
Both effects greatly enhance the \hmion\ continuum opacity, 
while the fraction of \feii\ to the total
number of iron atoms is unaffected since \feii\ is by far 
the dominant species.
As a consequence, the mean \fehii\ abundance of M55 appears
to be enhanced at given EWs.

In Figure~\ref{fig:photorelation} (l), we show a plot of \teff\ versus $\log g$ 
with a single relation, where RGB stars in both clusters are aligned well
on a single locus.
Figure~\ref{fig:photorelation} (j) and (k) show \dfeh\ against
\fehi\ and \fehii\ showing large discrepancies in M55 RGB stars,
reminiscent of the M22 \caw\ stars from Method 2 of Mu15
as shown in Figure~\ref{fig:method2}.
The metallicity distributions of \fehi\ and \fehii\ from 
a single \tefflogg\ relation shown in 
Figure~\ref{fig:m55n6752comp1} (b) and (c) are intriguing, since
the two peaks in the \fehii\ distribution become less conspicuous
with a single relation with respect to NGC~6752.

Our result with a mock GC strongly suggests that applying a single photometric
relation in order to derive the effective temperatures and surface gravities of 
individual stars in peculiar GCs with heterogeneous metallicities and 
perhaps ages, such as M22, may result in slightly incorrect 
metallicity scales and distributions.

\subsection{Spectroscopic method}\label{chap:spec}
We also performed a traditional analysis using the spectroscopic
effective temperatures and the surface gravities, which requires the excitation
and ionization equilibria of iron abundances.
Our results are shown in Figures~\ref{fig:m55n6752comp1} (i -- l) and 
\ref{fig:m55n6752comp2} (i -- l).
As shown in Table~\ref{tab:mock}, our mean spectroscopic \feh\ values are 
in good agreement with those of \citet{carretta09uves} and \citet{yong13}.
The difference in metallicity between M55 and NGC~6752 is
\dfehii\ = 0.32 -- 0.42 dex, depending on the data sets.
It should be mentioned that our spectroscopic \feh\ value of NGC~6752 
from \citet{carretta09uves} 
is about 0.1 dex higher than that from \citet{yong13}, which is consistent with 
the iron abundances of \feh\ = $-$1.56 dex \citep{carretta09uves} and 
$-$1.65 dex \citep{yong13} for the cluster.
The origin of this discrepancy of the mean metallicity of NGC~6752 is beyond
the scope of this study and we decline to discuss this matter further.

\subsection{Using the evolutionary $\log g$ with the spectroscopic \teff}\label{chap:fixedT}
\begin{figure}
\epsscale{1}
\figurenum{2}
\plotone{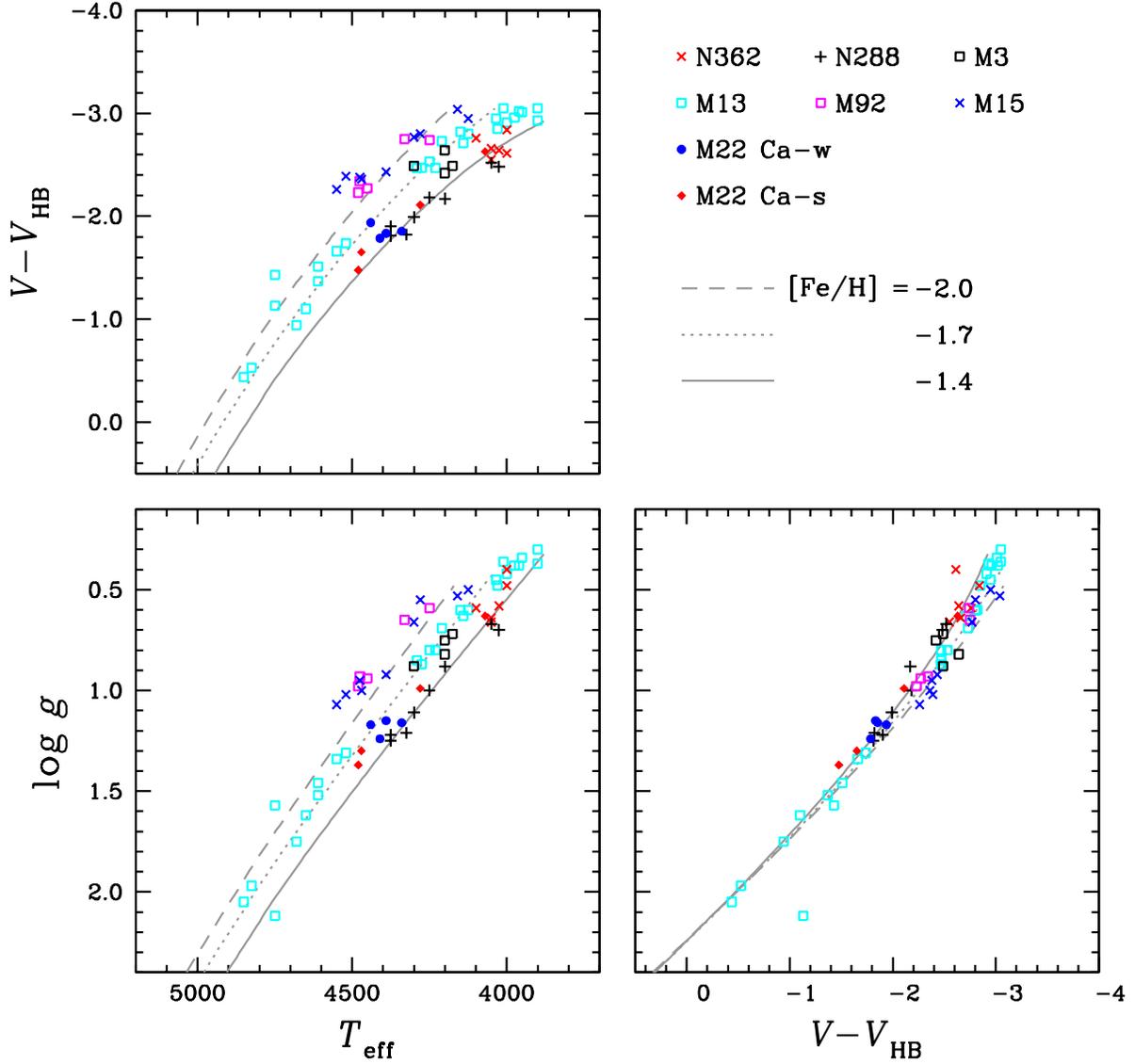}
\caption{Plots of $\log g$ versus \teff, \teff\ versus \vvhb,
and $\log g$ versus \vvhb\ for six GCs by \citet{ki03}.
Note the rather tight relation in $\log g$ versus \vvhb, suggesting
that the \vvhb\ magnitude can be a reddening- and distance-independent 
surface gravity indicator for metal-poor GC RGB stars.
Also shown are the Victoria-Regina model isochrones for 12 Gyr \citep{vr}.
}\label{fig:fit}
\end{figure}

\begin{figure}
\epsscale{1}
\figurenum{3}
\plotone{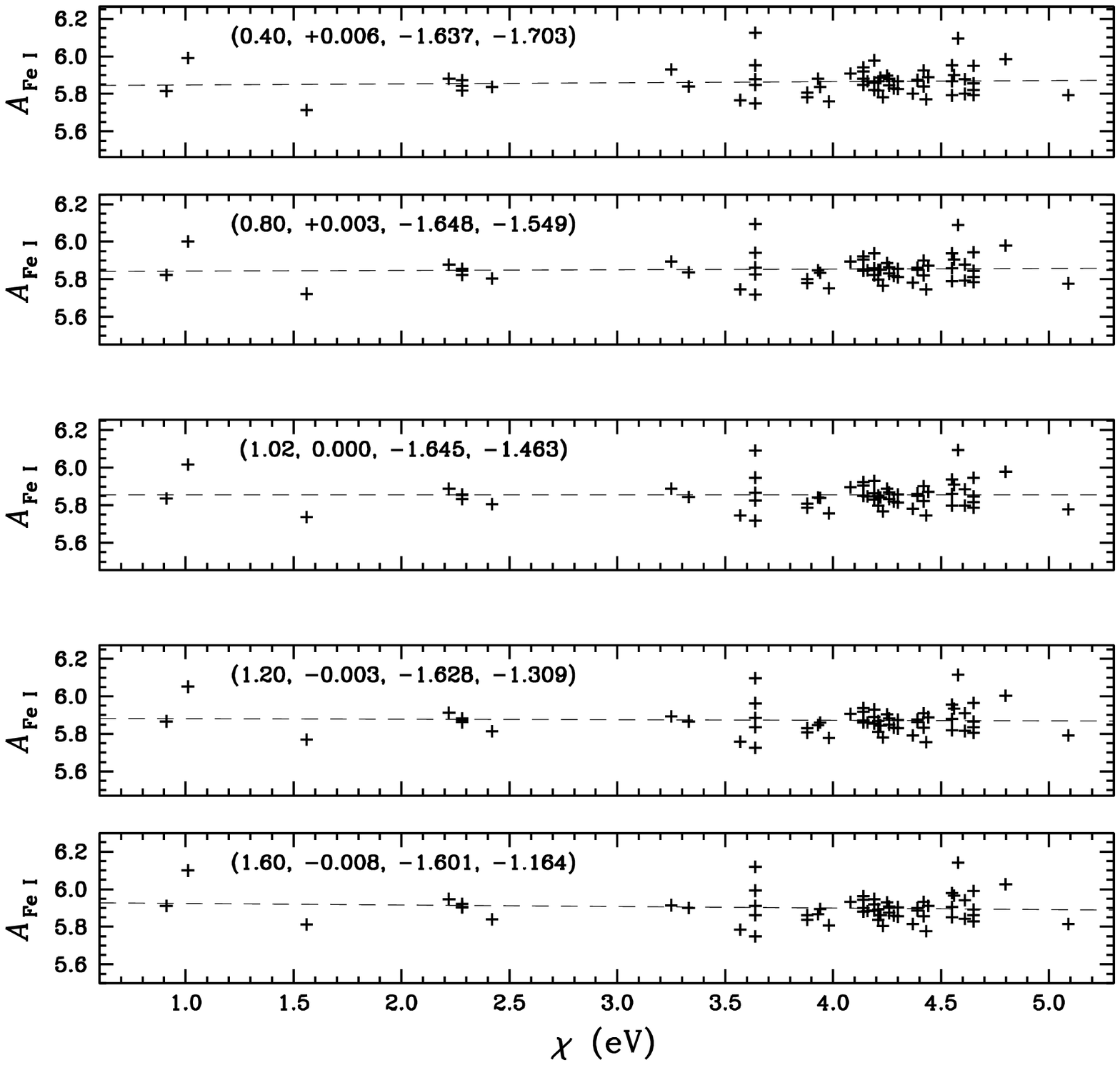}
\caption{Plots of the \afei\ against the excitation potential
for NGC6752-mg10 \citep{yong13}, adopting \teff\ = 4275 K.
Thin dashed lines indicate linear fits to the data.
The numbers in parentheses are the $\log g$ value in the cgs unit,
the slope in the excitation potential versus the iron abundance, 
\fehi\ and \fehii. 
Once the spectroscopic effective temperature is correctly determined,
the assumption of the excitation equilibrium holds for wide range of
the surface gravity.
}\label{fig:log}
\end{figure}

\begin{figure}
\epsscale{1}
\figurenum{4}
\plotone{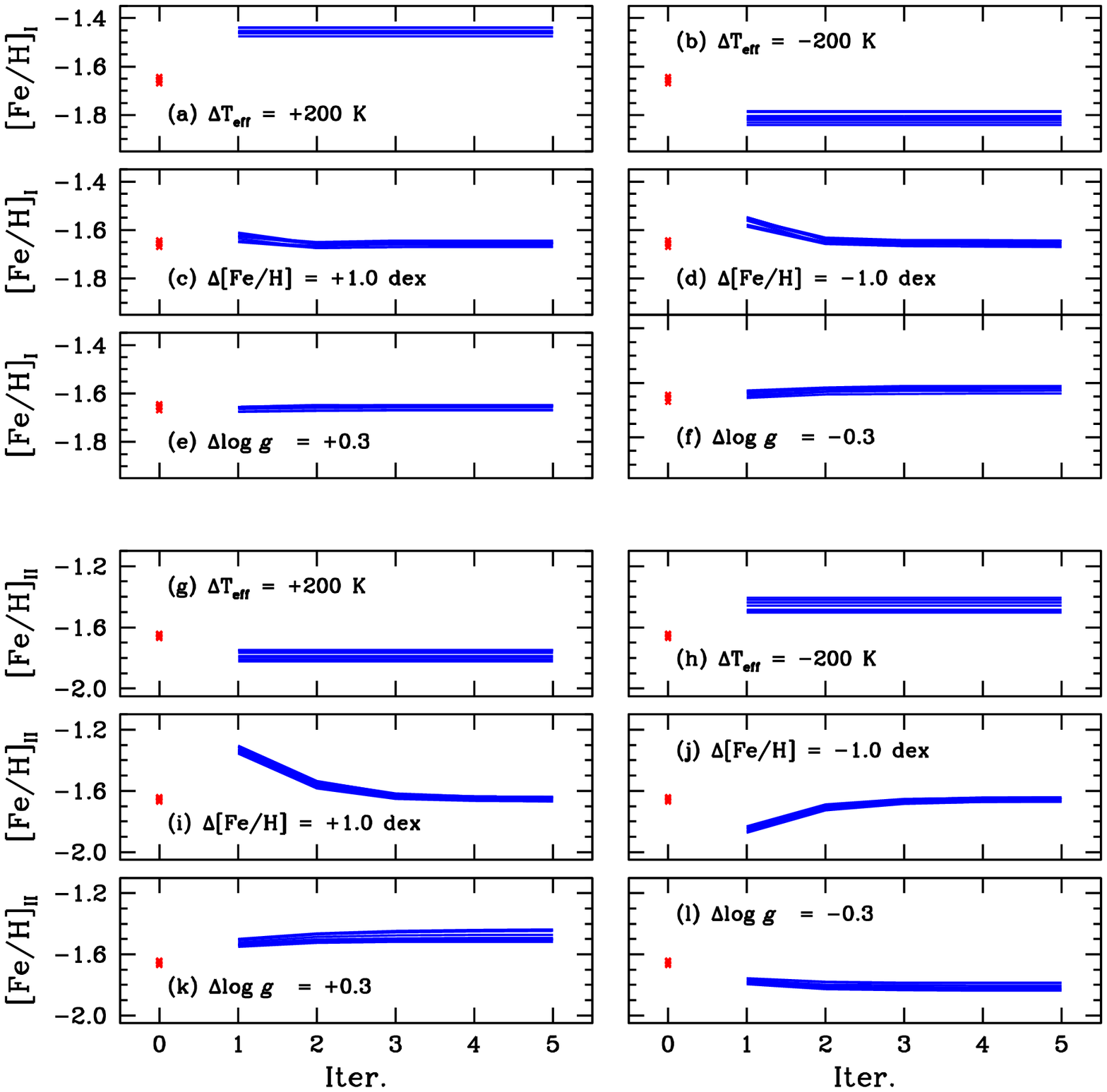}
\caption{
Iterative derivations of \fehi\ and \fehii\ abundances of 9 RGB stars 
in NGC~6752 using the EW measurements by \citet{yong13}. 
In each panel, the red crosses denote our spectroscopic \fehi\ and \fehii\
and blue solid lines are inferred abundances with iterations.
(a) and (b) \fehi\ abundances using input model atmospheres with
the effective temperature offsets of $\Delta$\teff\ = $\pm$200 K.
(c) and (d) Same as (a) and (b) but the initial metallicity offsets of $\Delta$\feh\ = $\pm$1.0.
(e) and (f) Same as (a) and (b) but the surface gravity offsets of  $\Delta\log g$ = $\pm$0.3.
(g) and (h) Same as (a) and (b) but for \fehii.
(i) and (j) Same as (c) and (d) but for \fehii.
(k) and (l) Same as (e) and (f) but for \fehii.
}\label{fig:fehvspam}
\end{figure}

\begin{figure}
\epsscale{1}
\figurenum{5}
\plotone{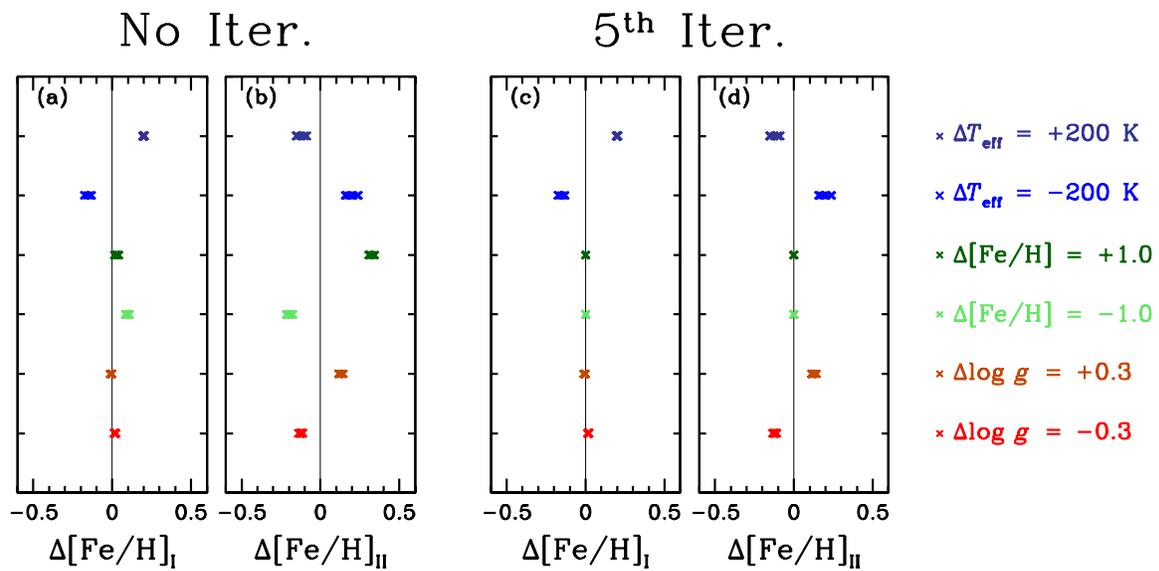}
\caption{
(a) and (b) The differences in \fehi\ and \fehii\ 
between those without iteration and those from spectroscopic method
for NGC~6752 RGB stars.
(c) and (d) Same as (a) and (b), but for the fifth iteration.
}\label{fig:delfeh}
\end{figure}

\begin{figure}
\epsscale{1}
\figurenum{6}
\plotone{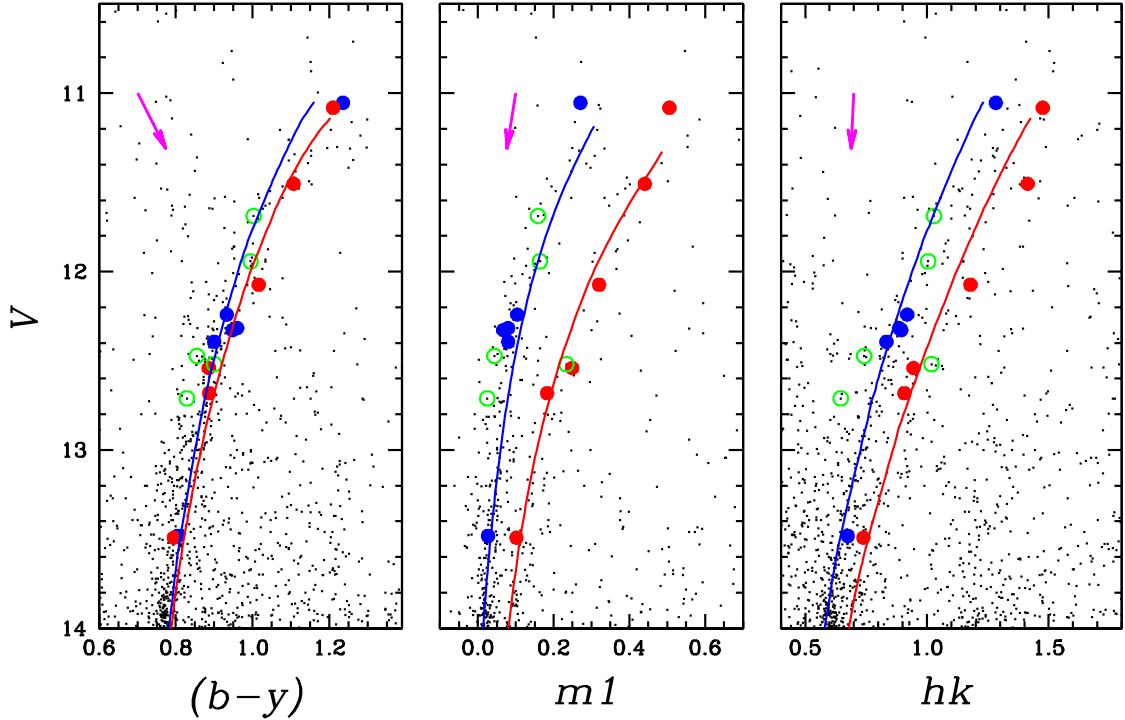}
\caption{Color-magnitude diagrams of M22 \citep{jwlnat,lee15}.
The filled blue (\caw) and red (\cas) circles denote RGB stars and
open green circles AGB stars tagged by Mu15.
The model isochrones for $(b-y)$ versus $V$ and $hk$ versus $V$ CMDs are 
from \citet{joo13}. The blue lines are for G1 (\feh\ = $-$1.96, $Y$ = 0.231,
12.8 Gyr) and the red lines are for G2 (\feh\ = $-$1.71, $Y$ = 0.32, 12.5 Gyr).
For the $m1$ versus $V$ CMD, we use model isochrones from \citet{vr} for
12 Gyr with [$\alpha$/Fe] = +0.3.
The blue line is for \feh\ = $-$1.53 and the red line is for \feh\ = $-$1.84.
Note that $m1$ index depends not only on overall metallicities but also
on lighter elemental abundances, such as CN.
The magenta arrows in each panel show reddening vectors corresponding 
to \ebv\ = 0.1 mag, and the differential reddening can not explain
the double RGB sequences in the $m1$ and the $hk$ CMDs.
}\label{fig:m22cmd}
\end{figure}

\begin{figure}
\epsscale{1}
\figurenum{7}
\plotone{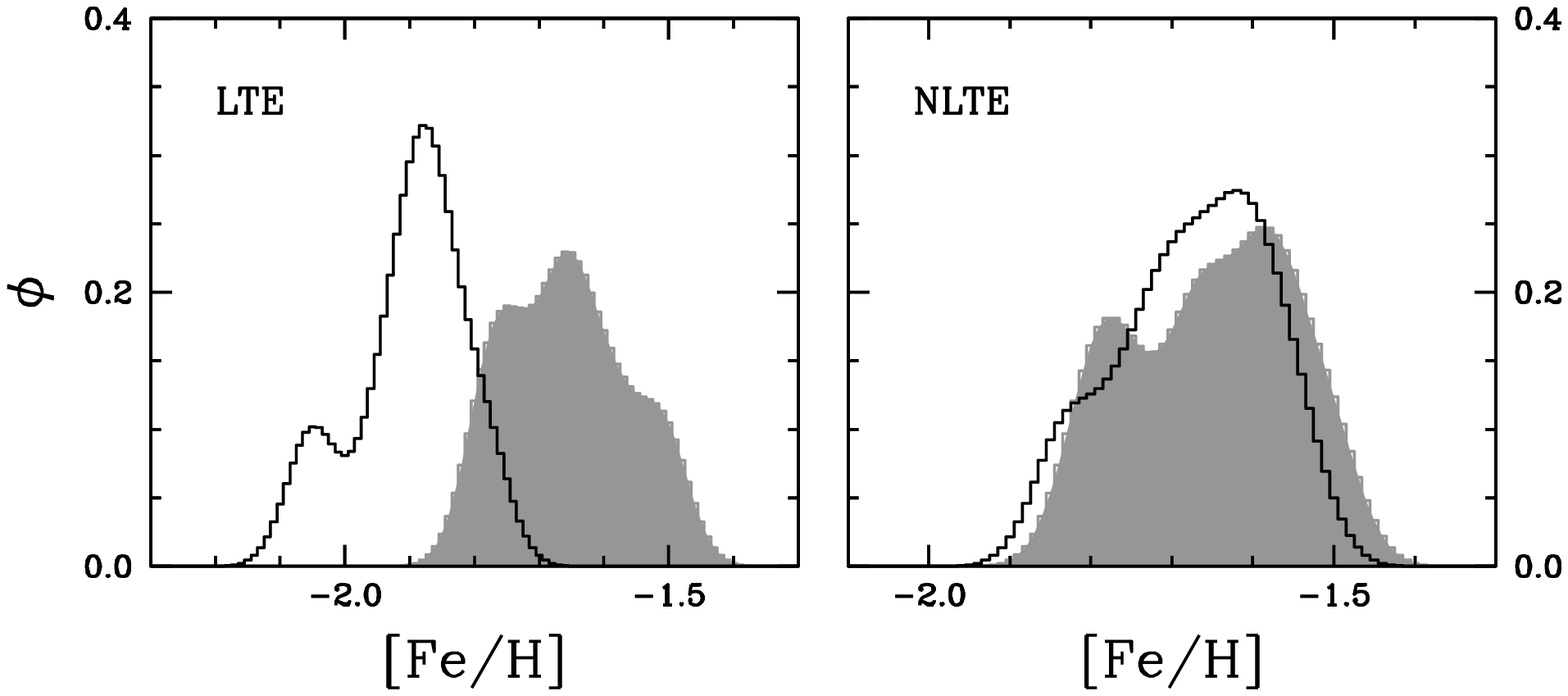}
\caption{ Metallicity distributions of M22 HB stars by \citet{marino13}.
The solid lines are for \fehi\ and the shades are for \fehii.
}\label{fig:hb}
\end{figure}

\begin{figure}
\epsscale{1}
\figurenum{8}
\plotone{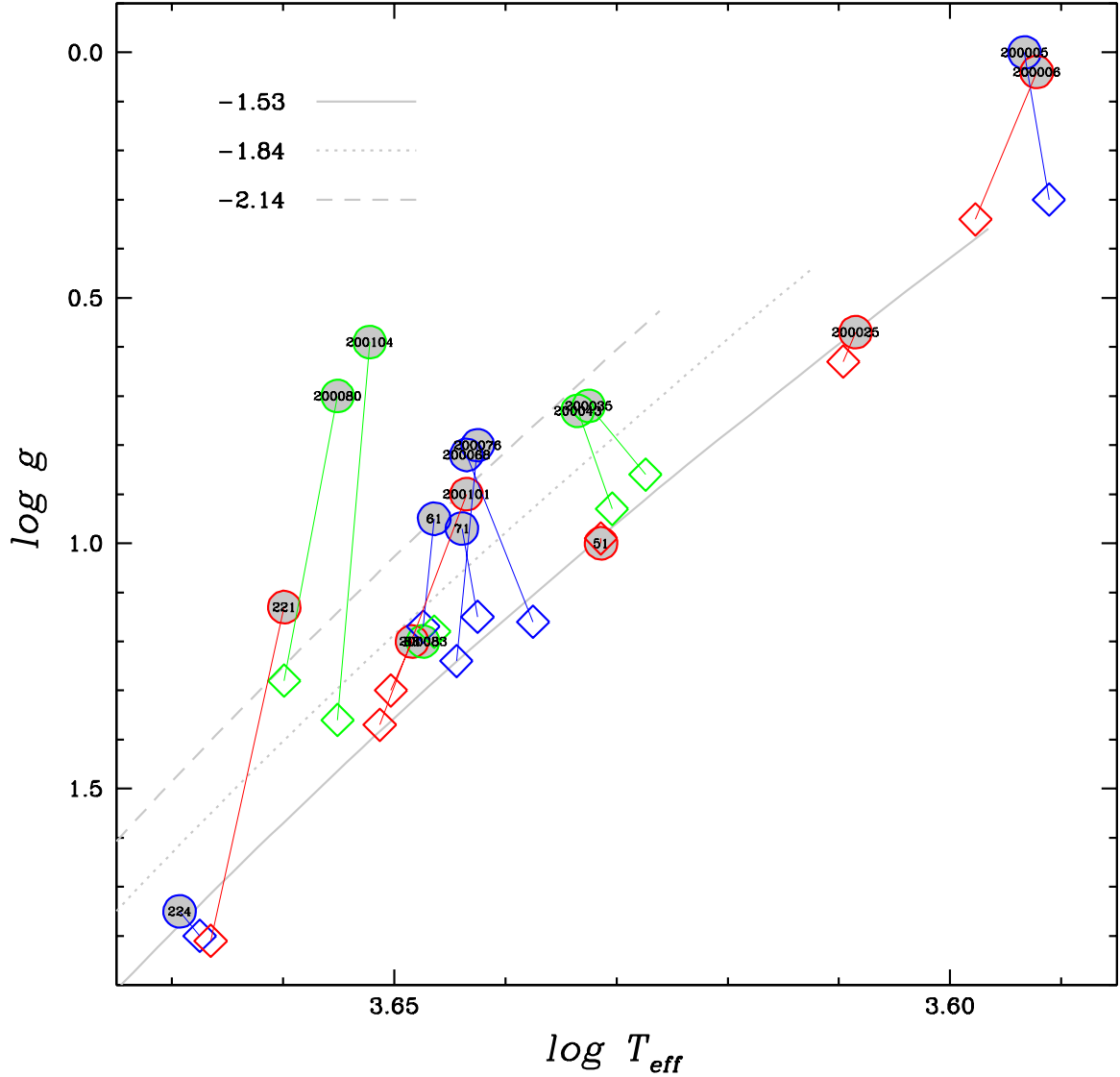}
\caption{Differences in $\log g$ and \teff\ between those from the Method 1
(filled circles; spectroscopic \teff\ and $\log g$) and 
the Method 2 (open diamonds; spectroscopic \teff\ and photometric $\log g$) 
of M22 spectroscopy target stars by Mu15.
Blue and red colors denote \caw\ and \cas\ RGB stars, respectively, and 
green color denotes AGB stars classified by Mu15. 
Note that the surface gravity of the \caw\ RGB stars are greatly
increased in the case of Method 2, i.e.\ the surface gravities
of \caw\ RGB stars from the Method 2 are preferentially larger than those from
the Method 1, which may lead increasing in the iron abundances 
from \ion{Fe}{2} lines for \caw\ RGB stars.
Also shown are the Victoria-Regina model isochrones for 12 Gyr \citep{vr}.
}\label{fig:pam}
\end{figure}

\begin{figure}
\epsscale{1}
\figurenum{9}
\plotone{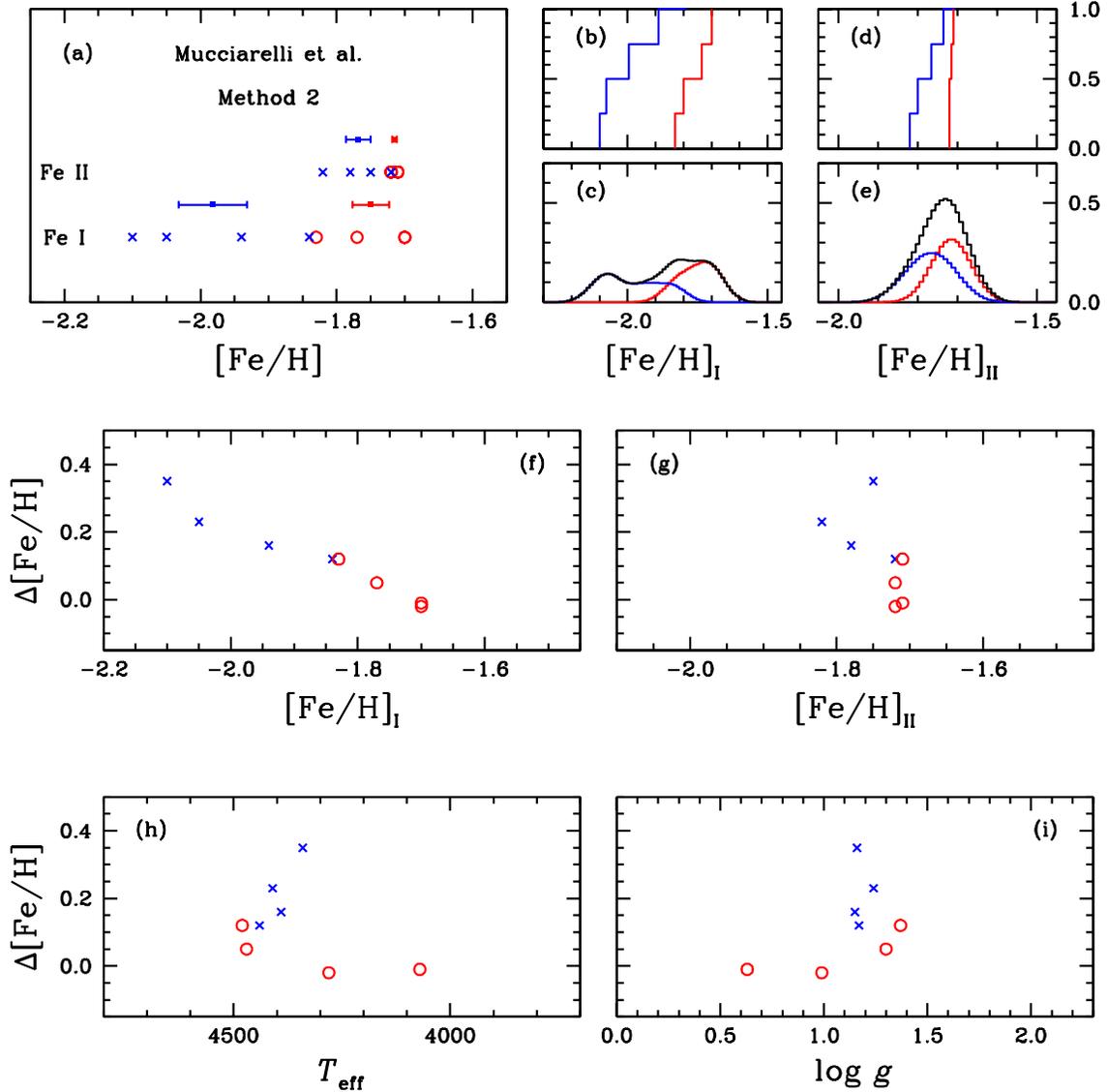}
\caption{
(a) Metallicity distributions of M22 RGB stars 
with 0.5 $\leq \log g \leq$ 1.5 (four stars in each group) from the Method 2 
of Mu15. The blue crosses are for the \caw\ RGB stars 
and the red circles the \cas\ RGB stars \citep{jwlnat,lee15}.
The horizontal bars indicate errors with a 2$\sigma$ range ($\pm$ 1$\sigma$).
The difference in the mean iron abundances between the two groups
are larger than a 2.5 $\sigma$ level both in \fehi\ and \fehii.
(b) -- (e) Cumulative and generalized metallicity distributions.
The blue and the red solid lines are for the \caw\ and 
\cas\ stars, respectively.
(f) -- (i) $\Delta$ [Fe/H] ( = \fehii\ $-$ \fehi) against \fehi,
\fehii, effective temperature and surface gravity.
Note that the $\Delta$ [Fe/H] values of the \caw\ RGB stars are preferentially
larger than those of the \cas\ RGB stars.
}\label{fig:method2}
\end{figure}

\begin{figure}
\epsscale{1}
\figurenum{10}
\plotone{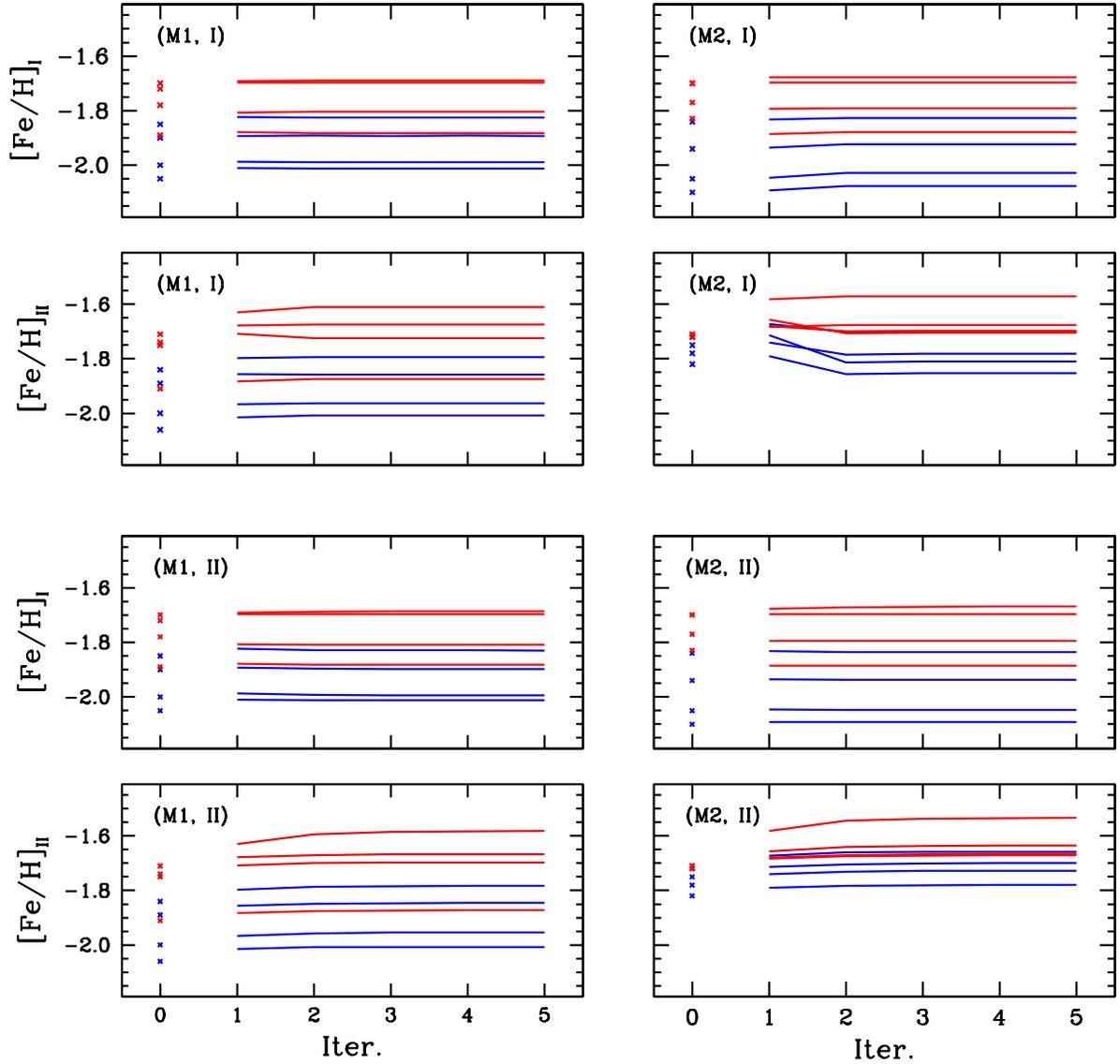}
\caption{
Iterative derivations of \fehi\ and \fehii\ of eight M22 RGB stars.
In each panel, the blue and the red color are for the \caw\ and the \cas\ 
RGB stars, respectively.
Crosses denote the metallicities of individual stars by Mu15
and solid lines denote inferred abundances with iterations.
M1 and M2 are for the metallicity derived using stellar parameters from
Method 1 and Method 2 of Mu15.
I and II refer to the reference metallicities, \fehi\ and \fehii\ respectively,
used in the calculations of the model atmospheres.
Note that the discrepancy in \fehii\ for \cas\ RGB stars from Method 2
are preferentially larger, indicating that the surface gravities 
of the \cas\ RGB stars adopted by Mu15 is most likely underestimated.
}\label{fig:Mu15iter}
\end{figure}

\begin{figure}
\epsscale{1}
\figurenum{11}
\plotone{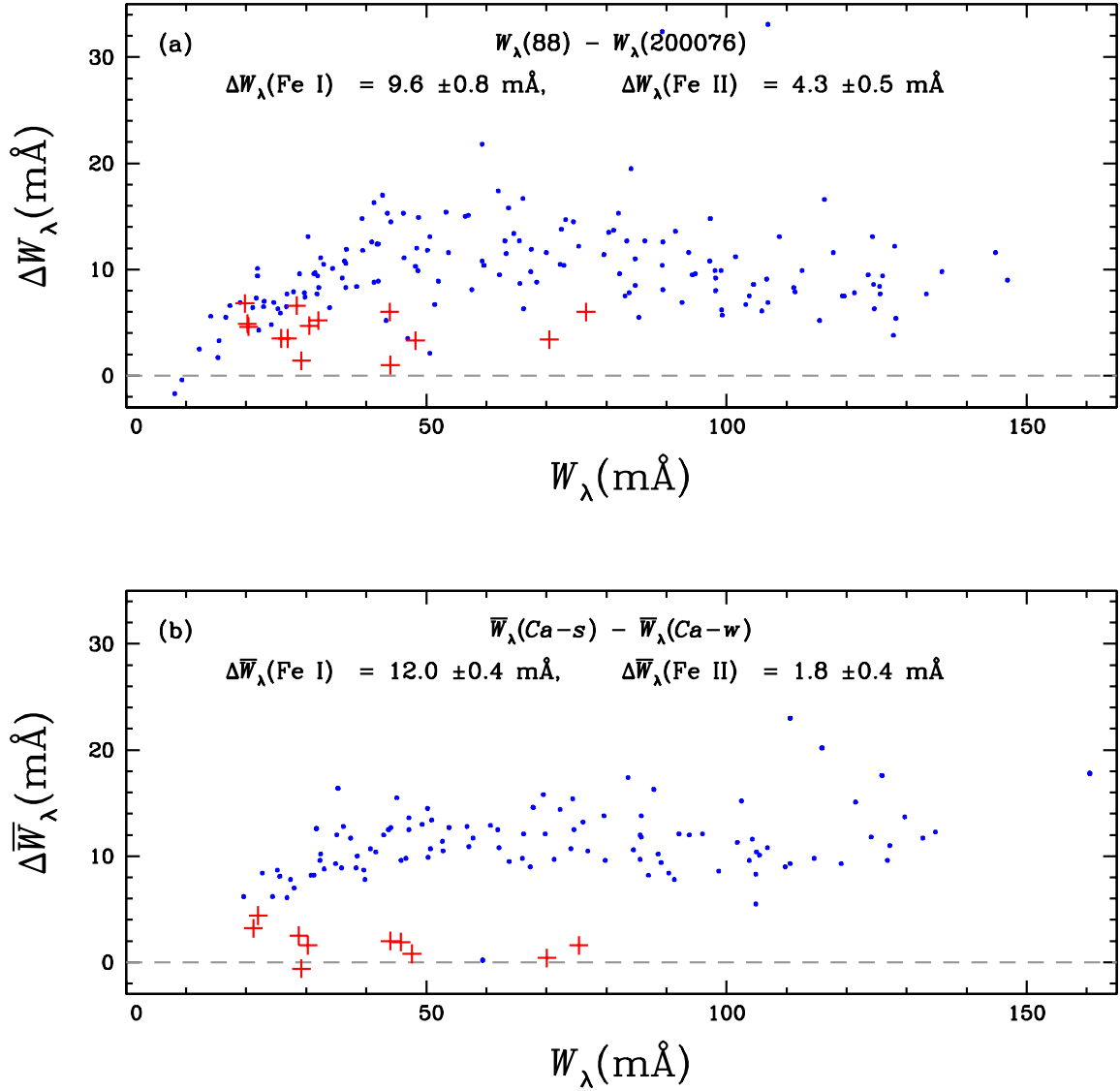}
\caption{(a) Differences in the equivalent widths between the \caw\ and 
the \cas\ RGB stars with similar visual magnitudes and colors,
EW(200076) $-$ EW(88). The blue dots are for the \ion{Fe}{1} lines
and the red plus signs are for the \ion{Fe}{2} lines.
In spite of similar visual magnitudes and colors between two RGB pairs,
the equivalent widths of the \cas\ RGB star (88) are stronger
than those of \caw\ star (200076), strongly indicate that
the \caw\ stars are more metal-poor than the \cas\ stars.
(b) Same as (a), but those of the mean values of each group.
}\label{fig:ew}
\end{figure}

\begin{figure}
\epsscale{1}
\figurenum{12}
\plotone{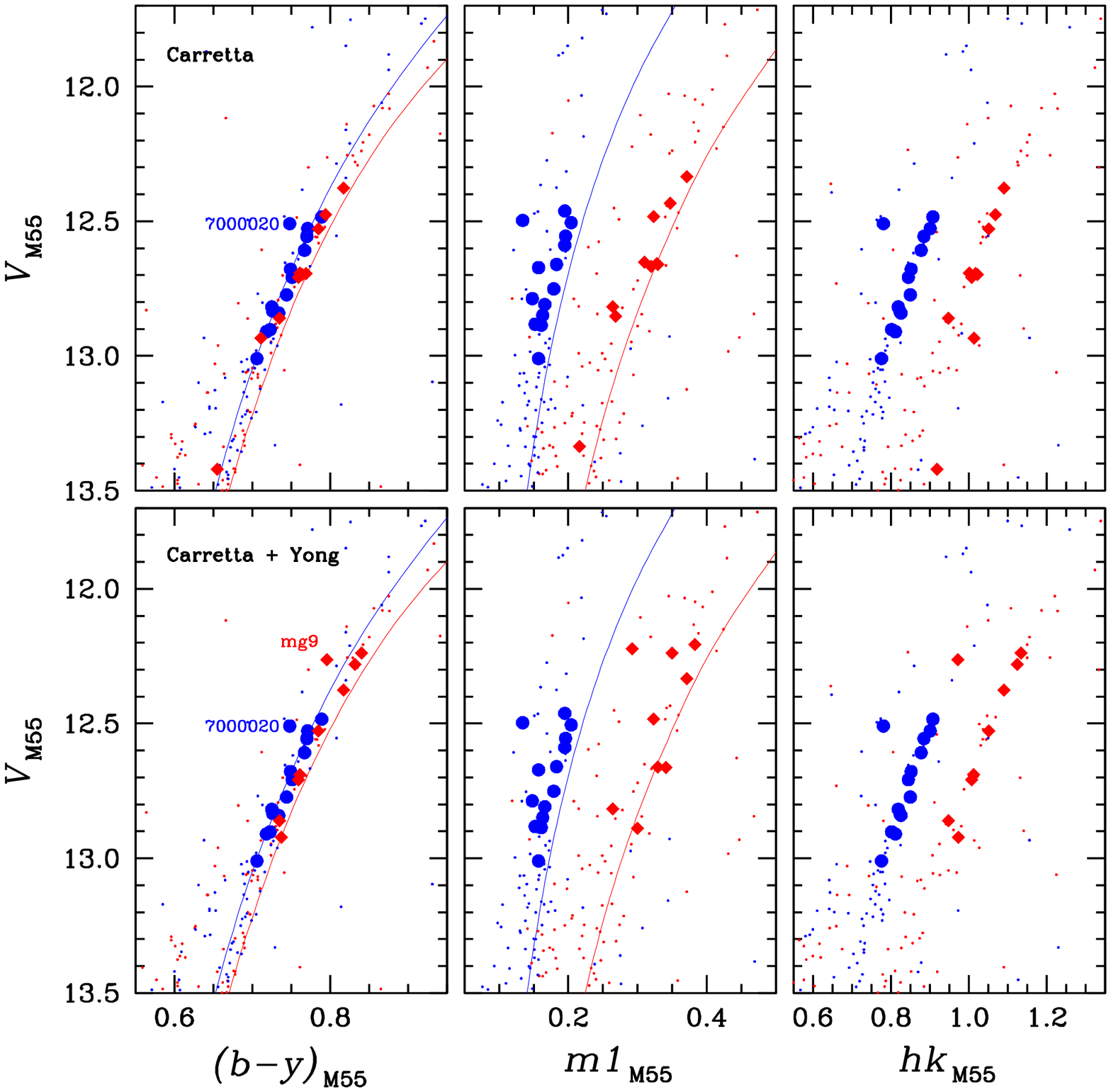}
\caption{A composite color-magnitude diagram for M55 (blue) and NGC~6752 (red).
(Upper panels) The filled blue circles and the filled red diamonds are M55 and NGC~6752
RGB stars studied by \citet{carretta09uves}, respectively.
Also shown are model isochrones for 12 Gyr with \feh\ = $-$1.84 (blue lines)
and $-$1.53 (red lines).
(Lower panels) Same as the upper panel but NGC~6752 RGB stars by \citet{yong13}.
Note that M55-7000020 and NGC~6752-mg9 appear to be AGB stars.
}\label{fig:m55n6752cmd}
\end{figure}

\begin{figure}
\epsscale{1}
\figurenum{13}
\plotone{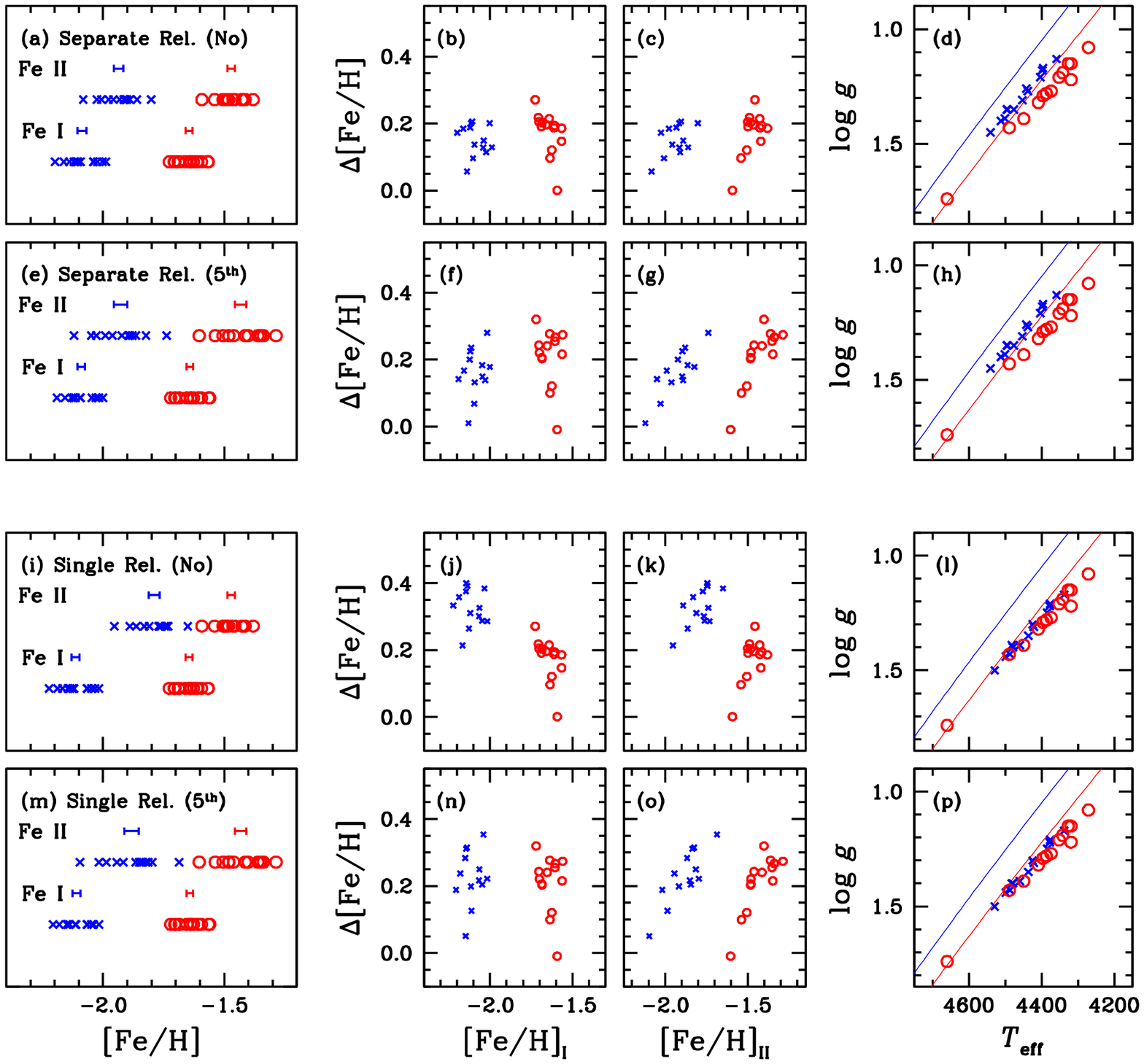}
\caption{
(a) Metallicity distributions of M55 and NGC~6752 with photometric
stellar parameters derived from two separate relations using 
\feh\ = $-$1.90 and $-$1.55 for M55 and NGC~6752, respectively.
(b) - (c) \dfeh\ against \fehi\ and \fehii. Note that the \dfeh\ ranges of 
both clusters agree well with those of other GCs in Figure~\ref{fig:ki03}.
(d) A plot of \teff\ and $\log g$ along with model isochrones
for 12 Gyr with \feh\ = $-$1.84 and $-$1.53 \citep{vr}.
(e) - (h) Same as (a) - (d) but those after the fifth iteration.
(i) - (l) Same as (a) - (d) but photometric stellar parameters 
derived from a single relation using \feh\ = $-$1.55 for both clusters.
Note that the \dfeh\ range of M55 is significantly larger than
that of NGC~6752 and those of other GCs in Figure~\ref{fig:ki03}.
Also note that the \dfeh\ range of M55 is comparable to that of the \caw\ stars
in M22 from Method 2 of Mu15 as shown 
in Figures~\ref{fig:ki03} and \ref{fig:method2}.
(e) - (h) Same as (a) - (d) but those after the fifth iteration, where
the discrepancy in \dfeh\ of M55 RGB stars becomes less severe 
with the iteration process.
}\label{fig:photorelation}
\end{figure}

As shown in Figure~\ref{fig:fit}, the model isochrones can provide a
useful means to derive the stellar parameters.
In Figure~\ref{fig:17gc}, we show the similar plots for 17 GCs 
from the homogeneous elemental abundance study by \citet{carretta09uves}.
Also shown are the Victoria-Regina isochrones for the age of 12 Gyr
and they appear to be in excellent agreement with observations.

We devise a new strategy to derive evolutionary surface gravities of 
RGB stars in GCs under the assumption that the excitation equilibrium of 
\ion{Fe}{1} lines is applicable in such stellar atmospheres and, furthermore,
excitation equilibrium holds for rather wide range of the surface gravity.
As shown in Figure~\ref{fig:fit} and \ref{fig:17gc}, at a given effective 
temperature, the surface gravity increases with metallicity,
and, as a consequence, the metallicity, especially \fehii, without the proper 
estimates of the surface gravity may not be correct.

\begin{figure}
\epsscale{1}
\figurenum{14}
\plotone{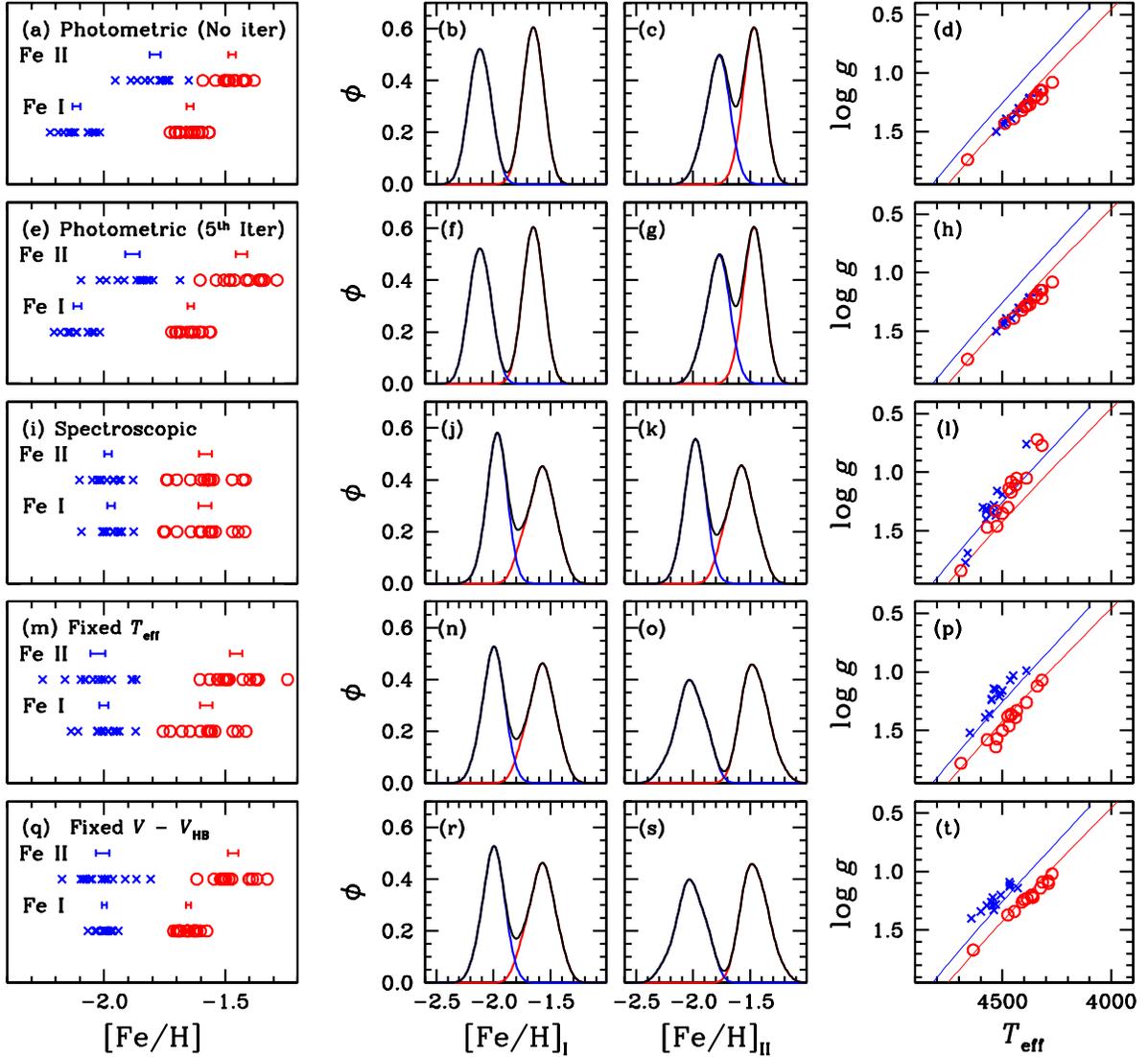}
\caption{Comparisons of [Fe/H] and stellar parameters between M55 and NGC~6752
RGB stars \citep{carretta09uves} from different approaches.
The blue color and the red color denote M55 and NGC~6752, respectively.
In plots of \teff\ versus $\log g$, we show model isochrones for 12 Gyr with 
\feh\ = $-$1.84 (blue lines) and $-$1.53 (red lines).
}\label{fig:m55n6752comp1}
\end{figure}

\begin{figure}
\epsscale{1}
\figurenum{15}
\plotone{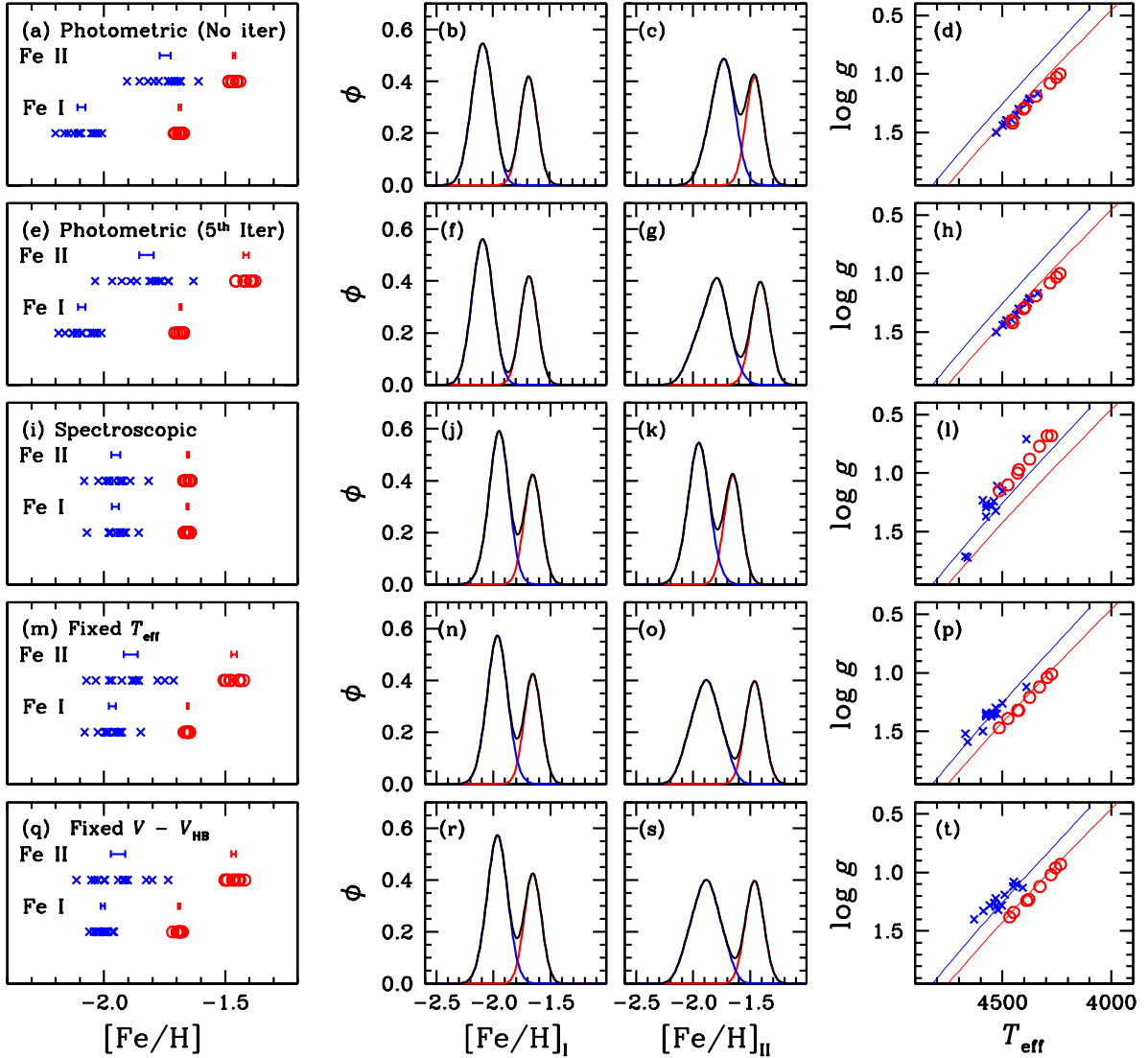}
\caption{Same as Figure~\ref{fig:m55n6752comp1} but using NGC~6752 RGB stars 
by \citet{yong13}.
The blue color and the red color denote M55 and NGC~6752, respectively.
Note that gf-values from \citet{yong13}, whose gf-values are slightly 
different from those adopted by \citet{carretta09uves}, used for both clusters 
and, as a consequence, the metallicity distributions of M55 
are slightly different from those in Figure~\ref{fig:m55n6752comp1}.
}\label{fig:m55n6752comp2}
\end{figure}

\begin{figure}
\epsscale{1}
\figurenum{16}
\plotone{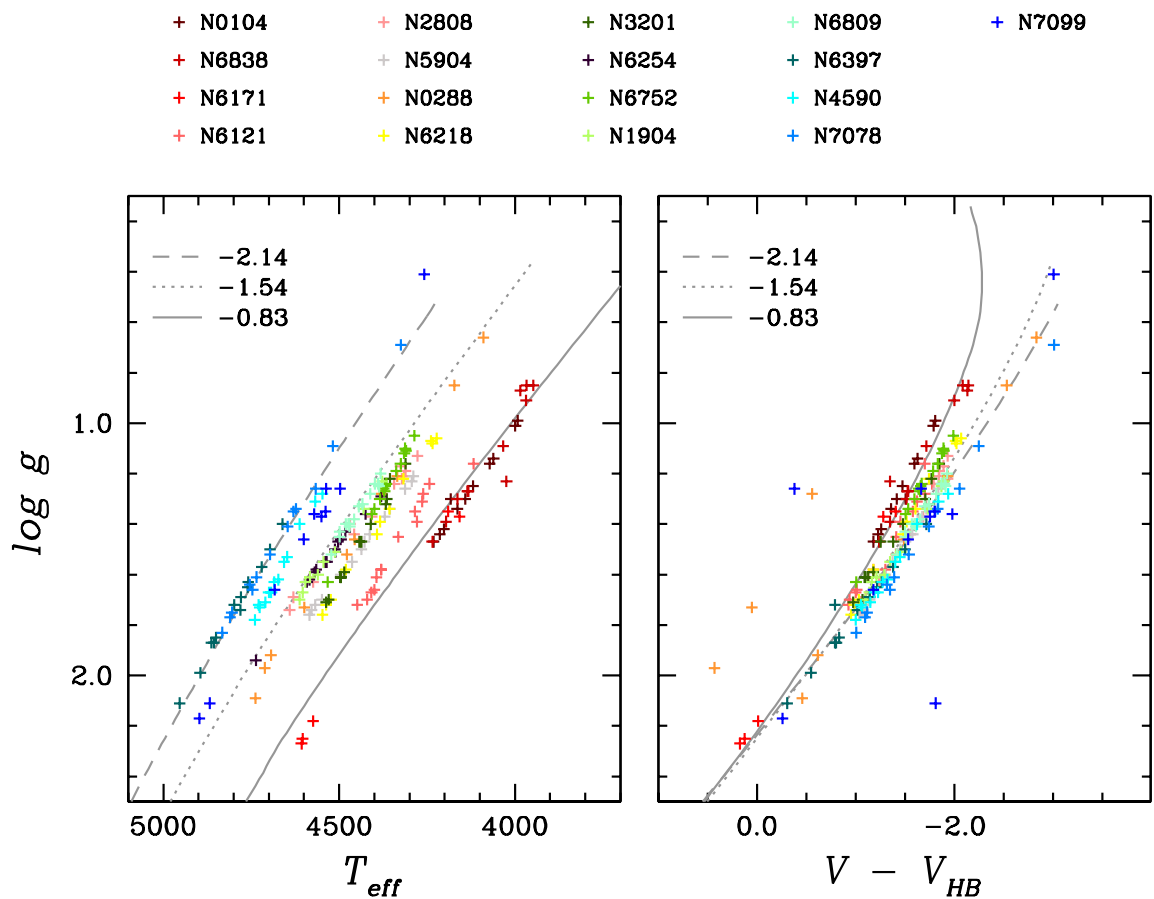}
\caption{Plots of $\log g$ versus \teff\ and $\log g$ versus \vvhb\ of RGB stars 
in 17 GCs by \citet{carretta09uves}.
Note the rather tight relation in $\log g$ versus \vvhb, similar to that 
in Figure~\ref{fig:fit}.
Also shown are the Victoria-Regina isochrones for 12 Gyr \citep{vr}.
}\label{fig:17gc}
\end{figure}

Using the spectroscopic temperature and the \fehii\ abundance from photometric 
method as initial input parameters, 
we interpolated the Victoria-Regina model isochrones to obtain 
the evolutionary surface gravity at the fixed effective temperature.
Then we derive the updated metallicity by running MOOG 
using the model atmosphere with the spectroscopic effective temperature 
and  the evolutionary surface gravity in an iterative manner 
until the derived metallicity converged to within
the internal measurement error between consecutive measurements,
which usually requires 2 to 3 iterations.
We show our new stellar parameters in
Figures~\ref{fig:m55n6752comp1} (p) and \ref{fig:m55n6752comp2} (p)
and metallicity distributions in Figures~\ref{fig:m55n6752comp1} (n -- o) and 
\ref{fig:m55n6752comp2} (n -- o).
The difference in metallicity between M55 and NGC~6752 becomes
$\Delta$\fehii\ = 0.44 -- 0.56 dex and our results are shown 
in Table~\ref{tab:mock}.
We also note that using the photometric temperatures of individual stars
does not change our results presented here.
This approach should be reddening- and distance-independent 
and therefore it would be useful to derive surface gravity of GC stars
with varying foreground reddening, such as RGB stars in M22.

\subsection{Using the evolutionary $\log g$ and \teff\ at a given \vvhb}
\label{chap:fixedVVHB}
As shown in Figure~\ref{fig:fit} and \ref{fig:17gc}, at a given metallicity,
the $V$ magnitude differences from the HB, \vvhb, of individual stars 
in GCs are well correlated with the surface gravities 
and the effective temperatures.
Similar to the previous approach, at a given \vvhb\ and metallicity 
we determine the evolutionary surface gravity and effective temperature 
simultaneously by interpolating the Victoria-Regina model isochrones.
For this purpose, we use our own photometry of the clusters 
and \fehii\ derived from the photometric stellar parameters as an initial guess
as have done previously. Then we derive the updated metallicity
by running MOOG using the model atmosphere with the evolutionary effective 
temperature and the surface gravity
in an iterative manner until the derived metallicity converged
to within the internal measurement error between consecutive measurements.
We show our results in Figures~\ref{fig:m55n6752comp1} (q -- t) and 
\ref{fig:m55n6752comp2} (q -- t). 
This approach provides similar results as those from 
the spectroscopic method and the method relying on the evolutionary surface
gravity.
The difference in metallicity between M55 and NGC~6752 becomes
$\Delta$\fehii\ = 0.44 -- 0.52 dex as shown in Table~\ref{tab:mock}.
The merit of using \vvhb\ is that it is also a reddening- and 
distance-independent parameter.
However it can be vulnerable to the differential foreground reddening effect
of the individual stars.

\bigskip
It should be kept in mind that the main idea to deliver in our study is 
to demonstrate the importance of having appropriate stellar parameters 
for the LTE abundance analysis in multiple stellar populations.

\section{Revisiting the metallicity spread in M22}
Following the same procedures as for M55 and NGC~6752,
we derive the iron abundances of the two groups of stars in M22
in four different manners.
Our results are consistent with the idea that the two groups of stars
in M22 have different mean iron abundances as \citet{jwlnat}, \citet{lee15}
and \citet{marino09,marino11} already showed.

\subsection{Photometric method using a single relation}
\begin{figure}
\epsscale{1}
\figurenum{17}
\plotone{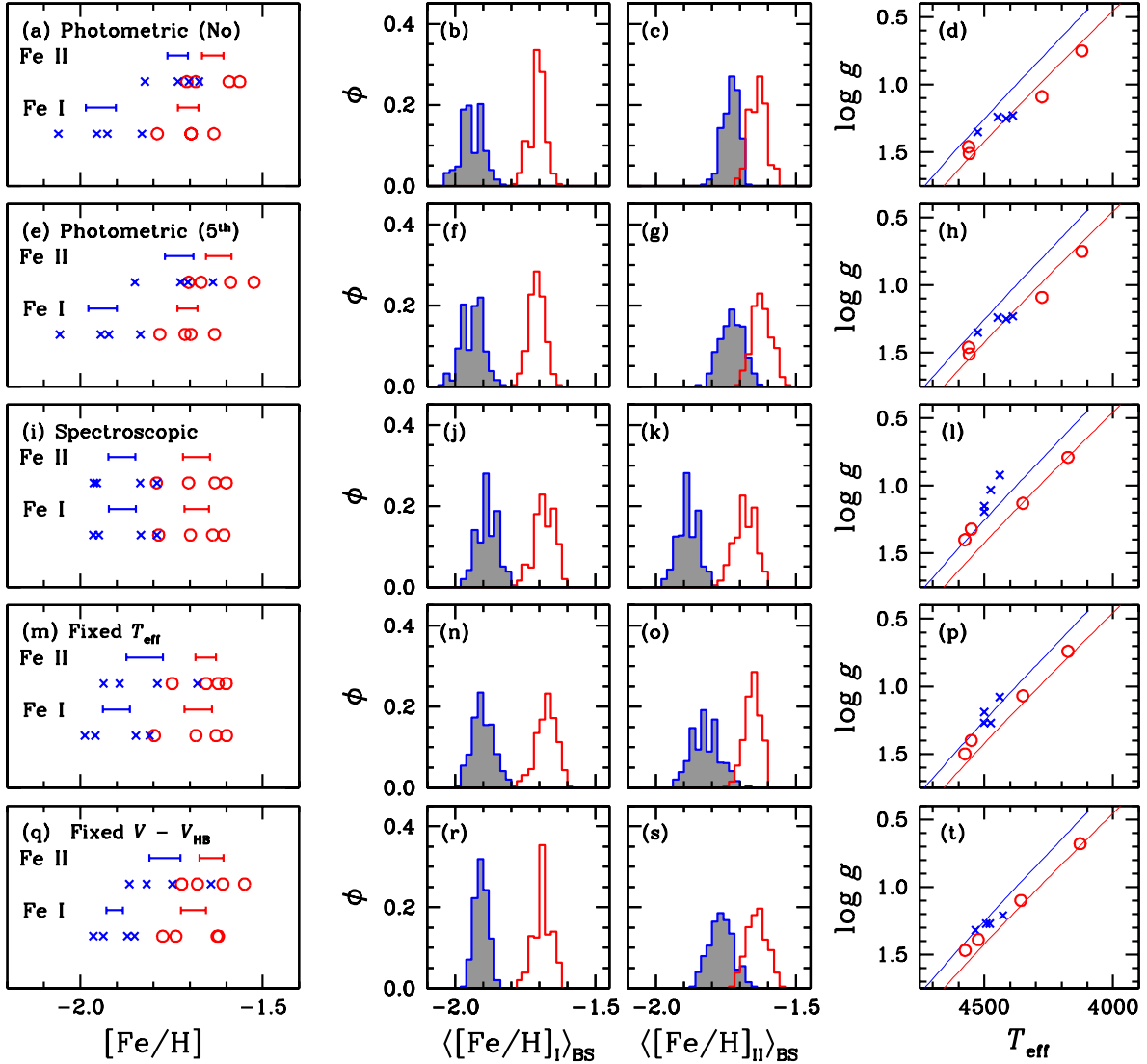}
\caption{Same as Figures~\ref{fig:m55n6752comp1} and \ref{fig:m55n6752comp2}
but for M22. 
The blue color and the red color denote the \caw\ and the \cas\ groups, 
respectively.
The histograms in the middle panels show empirical distributions of 
the mean \fehi\ and the mean \fehii\ for both populations from
the bootstrap method, strongly suggest  that the metallicity distributions 
of the \caw\ and the \cas\ groups are not identical.
}\label{fig:m22comb}
\end{figure}

\begin{figure}
\epsscale{1}
\figurenum{18}
\plotone{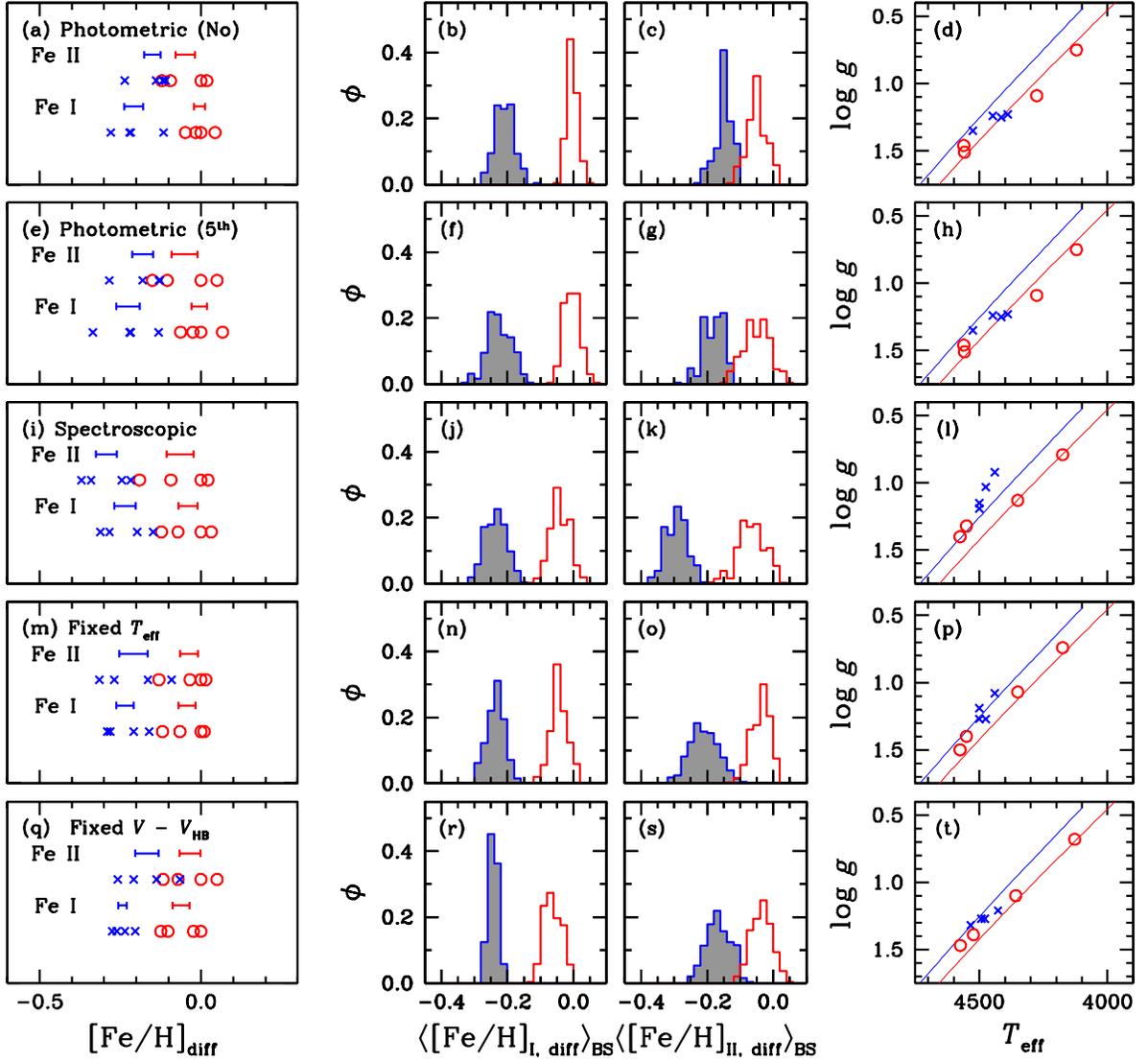}
\caption{Same as Figure~\ref{fig:m22comb}
but for differential analysis with respect to the star 51.
}\label{fig:m22combdiff}
\end{figure}

We derive the metallicity of M22 RGB stars based on the photometric 
effective temperature and surface gravity from our \str\ photometry 
of the cluster using the relations by \citet{alonso99}.
During our calculations, we adopted the apparent visual distance modulus 
of 13.60 mag, \ebv\ = 0.34 and \feh\ = $-$1.65 for M22 \citep{harris96}.
As it was done before, we made use of the weak lines only, \rew\ $\leq  -5.2$, 
for both \ion{Fe}{1} and \ion{Fe}{2} in order to minimize the effect 
of the adopted micro-turbulent velocity on the metallicity.
In Table~\ref{tab:ks} and Figures~\ref{fig:m22comb} and \ref{fig:m22combdiff},
we show our results.
For non-differential analysis, the differences in the mean metallicity are 
\dfehi\ = 0.239 $\pm$ 0.057 dex and \dfehii\ = 0.096 $\pm$ 0.048 dex
without iteration,
and \dfehi\ = 0.233 $\pm$ 0.048 dex and \dfehii\ = 0.108 $\pm$ 0.052 dex
after fifth iteration.
Note that our results are consistent with those from the Method 2 by Mu15.
In panels (b) and (c) of Figures~\ref{fig:m22comb} and \ref{fig:m22combdiff},
we show empirical distributions of the mean \fehi\ and \fehii\ for the two
populations from the bootstrap method, strongly suggest that 
the metallicity distributions of the two groups of stars in M22 are different.
As shown in Table~\ref{tab:ks},
the significance levels to reject the hypothesis that the mean \fehi\ values
of the \caw\ and the \cas\ groups are identical are lower than 1\%.
However, those for \fehii\ are rather large, $\approx$ 12 \%, 
for the non-differential analysis.
It should not be confused that this rather large significance levels do not
indicate that two groups of stars in M22 belong to the same population,
but the LTE analysis of the heterogeneous groups of star with a single
\tefflogg\ relation may be in error.

We also calculate the  line-by-line differential iron abundances since
the numbers of iron lines being measured by Mu15
for individual stars are different.
We selected the star 51 to be the reference star since its \teff\ and $\log g$
are close to the average for the sample. Also the stellar parameters for
this star both from the photometric and spectroscopic methods by 
Mu15 agree well as shown in Figure~\ref{fig:pam}.
For our differential abundance measurements, 
we did not adjust the stellar parameters of individual stars with respect
to the reference star as have done by \citet{yong13}, for example,
and we intended to calculate the proper  metallicity offset differences
among the sample stars with given stellar parameters.
The differences in the mean metallicity from the differential analysis are
\dfehi\ = 0.203 $\pm$ 0.039 dex and \dfehii\ = 0.101 $\pm$ 0.045 dex
without iteration,
and \dfehi\ = 0.220 $\pm$ 0.043 dex and \dfehii\ = 0.129 $\pm$ 0.051 dex
with fifth iterations,
consistent with those from non-differential analysis.
As shown in Table~\ref{tab:ks}, the separation of the mean \fehii\ values
between the \caw\ and the \cas\ groups is larger than 2.0 -- 2.5 $\sigma$ levels.
We calculated the significance levels to reject the hypothesis that the mean
\feh\ values of the two groups of stars in M22 are identical. 
We obtained the significance levels lower than 1\% for \fehi\ and 
7.5\% for \fehii,
strongly suggesting that they are different.

\subsection{Spectroscopic method}
Next, we derived the metallicity of individual stars based on the spectroscopic
\teff\ and $\log g$.
The differences in the mean metallicity between the two groups of stars
are \dfehi\ = 0.203 $\pm$ 0.050 dex and \dfehii\ = 0.204 $\pm$ 0.052 dex 
for non-differential analysis and
\dfehi\ = 0.194 $\pm$ 0.044 dex and \dfehii\ = 0.228 $\pm$ 0.053 dex 
for differential analysis,
making the separation of the mean \fehii\ values between the two groups 
larger than 3.9 $\sigma$ to 4.3 $\sigma$ levels.
Not surprisingly, our results are consistent with those obtained 
by \citet{marino09,marino11}, who relied on the traditional spectroscopic
stellar parameters.
As shown in Table~\ref{tab:ks}, 
the significance levels to reject the hypothesis that the mean
\feh\ values of the two groups of stars in M22 are identical are very low,
indicating that they are different.

\subsection{Using the evolutionary $\log g$ with the spectroscopic \teff}
The metallicity based on the evolutionary stellar parameters
also suggest that the metallicity distributions of each group of stars
are indeed different.
Following the same procedure described in \S{\ref{chap:fixedT}},
we obtained the differences in the mean metallicity of
\dfehi\ = 0.224 $\pm$ 0.061 dex and \dfehii\ = 0.168 $\pm$ 0.066 dex
for non-differential analysis and
\dfehi\ = 0.191 $\pm$ 0.043 dex and \dfehii\ = 0.172 $\pm$ 0.060 dex 
for differential analysis.
The mean \fehii\ values of each group are different
more than 2.5 $\sigma$ to 2.9 $\sigma$ levels and the low 
significance levels of being identical distributions also
confirm that they are different.

\subsection{Using the evolutionary $\log g$ and \teff\ at a given \vvhb}
Similar conclusion can be drawn when we use
the evolutionary $\log g$ and \teff\ at a given \vvhb,
where we used \vhb\ = 14.15 mag for M22 \citep{harris96}.
Following the same procedure described in \S{\ref{chap:fixedVVHB}},
we obtained the differences in the mean metallicity of
\dfehi\ = 0.216 $\pm$ 0.047 dex and \dfehii\ = 0.128 $\pm$ 0.062 dex
for non-differential analysis and
\dfehi\ = 0.181 $\pm$ 0.034 dex and \dfehii\ = 0.132 $\pm$ 0.056 dex 
for differential analysis.
Similar to the results shown above, 
the mean \fehii\ values of each group are different
more than 2.1 $\sigma$ to 2.4 $\sigma$ levels and the low 
significance levels of being identical distributions also
confirm that they are different.

As shown in Figures~\ref{fig:m22comb} and \ref{fig:m22combdiff},
it should be emphasized that 
the substructures not only in the \fehi\ but also in the \fehii\ distributions 
are notable, indicating that M22 contains multiple stellar populations with
heterogeneous metallicities.

\section{SUMMARY}
The precision elemental abundance measurement of individual stars in GGs 
is not a trivial task, especially for a peculiar GC with multiple stellar
populations with heterogeneous metallicity distributions.
In the context of LTE analysis,
our demonstrations with a mock peculiar GC composed of the two normal GCs, 
NGC~6752 and M55, showed that the internal absolute and the relative 
metallicity scales are vulnerable to the incorrect treatment of the input stellar 
parameters among multiple stellar populations with different metallicity. 
In particular, photometric surface gravity without taking care of proper 
metallicity effect can result in the \fehii\ measurement error of 
as large as 0.1 -- 0.2 dex.
As we have discussed earlier, this is because
the \feii\ line opacity does not vary with surface gravity since 
the almost all iron atoms are populated in the first ionized level,
while the \hmion\ continuum opacity is sensitively 
dependent on the electron pressure and, therefore, surface gravity. 
As a consequence, changes in surface gravity can mimic 
the \fehii\ abundance of RGB stars in GCs.
In this regard, we developed methods independent of the traditional 
spectroscopic analysis approach, which is demanding excitation and ionization 
equilibria in \ion{Fe}{1} and \ion{Fe}{2} elements, 
to make use of the evolutionary surface gravity. 
The metallicity scales from these new approaches are in good agreement 
with those of previous studies by others \citep{carretta09uves,yong13}.

From our study of a mock peculiar GC, it is worth to mention three comments 
concerning the metallicity scale of the multiple stellar populations in a GC.
First, the use of narrow band photometry, such as $m1$ and $hk$  which are 
sensitive to metallicity and less sensitive to interstellar reddening,
is beneficial to discern small \feh\ differences.
Second, our results show that our adaptive methods to estimate appropriate
surface gravity would be essential in deriving the absolute and the relative
\fehii\ scale for the multiple stellar populations in peculiar GCs.
Third, it is very interesting to note that our new methods appear to provide
similar metallicity scale as that from the traditional spectroscopic analysis,
suggesting that, the metallicity scale from the widely used 
traditional spectroscopic approach which makes use of excitation and 
ionization equilibria in \ion{Fe}{1} and \ion{Fe}{2} elements, is valid, 
at least in the relative sense.

Contrary to the conclusion made by Mu15, our re-examination 
of M22 RGB stars showed that the peculiar GC M22 is composed of 
two groups of stars with heterogeneous metallicities, confirming
our previous results and those of others 
\citep{jwlnat,lee15,dacosta09,marino09,marino11},
and the M22 saga will continue.

\acknowledgements
This work has been supported by the Center for Galaxy Evolution 
Research (grant no. 2010-0027910) and 
the Basic Science Research Program (grant no. 2016R1A2B4014741)
through the National Research Foundation of Korea (NRF) 
funded by the Korea government (MSIP).
J.-W. Lee thanks Eugenio Carretta for kindly providing EW measurements for 
NGC~6752 and M55 RGB stars and David Yong and Donghoh Kim for their helpful
discussions.
He also appreciates the anonymous referee 
for critical reading and for providing helpful suggestions
that greatly improved the manuscript.


\clearpage

\begin{deluxetable}{crrrrr}
\tablecaption{Mean abundance dependence on model atmosphere.\label{tab:dep}}
\tablenum{1}
\tablewidth{0pc}
\tablehead{
\multicolumn{1}{c}{} &
\multicolumn{2}{c}{$\delta$\teff} &
\multicolumn{1}{c}{} &
\multicolumn{2}{c}{$\delta\log g$}\\
\cline{2-3}\cline{5-6}
\multicolumn{1}{c}{} &
\multicolumn{1}{c}{+50 K} &
\multicolumn{1}{c}{$-$50 K} &
\multicolumn{1}{c}{}&
\multicolumn{1}{c}{+0.2} &
\multicolumn{1}{c}{$-$0.2}}
\startdata
\fehi  & 0.044 $\pm$ 0.003 & $-$0.044 $\pm$ 0.003 &&
                 $-$0.007 $\pm$ 0.002 & 0.000 $\pm$ 0.001 \\
\fehii  & $-$0.037 $\pm$ 0.006 & 0.036 $\pm$ 0.005 &&
                 0.090 $\pm$ 0.004 & $-$0.084 $\pm$ 0.004 \\
\enddata
\end{deluxetable}

\clearpage

\begin{landscape}
\begin{deluxetable}{crrrrrrrrrr}
\tablecaption{Differences in metallicity between two groups of stars
from different approaches.\label{tab:ks}}
\tabletypesize{\footnotesize}
\tablenum{2}
\tablewidth{0pc}
\tablehead{
\multicolumn{1}{c}{Method} &
\multicolumn{4}{c}{\fehi} &
\multicolumn{1}{c}{} &
\multicolumn{4}{c}{\fehii}\\
\cline{2-5}\cline{7-10}
\multicolumn{1}{c}{} &
\multicolumn{1}{c}{$\Delta$} &
\multicolumn{1}{c}{{\it t\tablenotemark{1}}} &
\multicolumn{1}{r}{S.L.\tablenotemark{2}}&
\multicolumn{1}{r}{S.L.\tablenotemark{3}}&
\multicolumn{1}{c}{} &
\multicolumn{1}{c}{$\Delta$} &
\multicolumn{1}{c}{{\it t\tablenotemark{1}}} &
\multicolumn{1}{r}{S.L.\tablenotemark{2}} &
\multicolumn{1}{r}{S.L.\tablenotemark{3}} \\
\multicolumn{1}{c}{} &
\multicolumn{1}{c}{} &
\multicolumn{1}{c}{} &
\multicolumn{1}{r}{(\%)}&
\multicolumn{1}{r}{(\%)}&
\multicolumn{1}{c}{} &
\multicolumn{1}{c}{} &
\multicolumn{1}{c}{} &
\multicolumn{1}{r}{(\%)} &
\multicolumn{1}{r}{(\%)}
}
\startdata
Mu15 M2     & 
0.232 $\pm$ 0.066 & 3.52 &  1.93 & 0.00 && 0.053 $\pm$ 0.022 & 2.43 &  8.98 & 0.00 \\

Mu15 M2 (5$^{\rm th}$ Iter.)    & 
0.217 $\pm$ 0.076 & 2.84 &  3.02 & 3.14 &&  0.090 $\pm$ 0.041 & 2.20 &  7.24 & 5.57 \\

\teff\ + $\log g$ (A99, No Iter.) & 
0.239 $\pm$ 0.057 & 4.20 &  0.76 & 0.00 && 0.096 $\pm$ 0.048 & 2.01 &  9.11 & 11.27 \\

\teff\ + $\log g$ (A99, 5$^{\rm th}$ Iter.) & 
0.233 $\pm$ 0.048 & 4.24 &  0.72 & 0.00 && 0.108 $\pm$ 0.052 & 1.80 &  12.32 & 11.42 \\

Spectroscopic & 
0.203 $\pm$ 0.050 & 3.48 &  1.32 & 0.00 && 0.204 $\pm$ 0.052 & 3.38 &  1.49 & 3.02 \\

Fixed \teff &
0.224 $\pm$ 0.061 & 3.67 &  1.04 & 0.00 && 0.168 $\pm$ 0.066 & 2.54 &  5.44 & 2.85 \\

Fixed \vvhb & 
0.216 $\pm$ 0.047 & 4.57 &  0.52 & 0.00 && 0.128 $\pm$ 0.062 & 2.08 &  8.56 & 8.46 \\

 &  &  &  & &&  &  &  & \\
\multicolumn{8}{c}{\it Differential Analysis} \\
Mu15 M2  & 
0.213 $\pm$ 0.064 & 3.30 &  1.92 & 0.00 && 0.077 $\pm$ 0.033 & 2.35 &  8.01 & 2.89 \\

Mu15 M2 (5$^{\rm th}$ Iter.)    & 
0.205 $\pm$ 0.069 & 2.96 &  2.60 & 2.71 &&  0.112 $\pm$ 0.036 & 3.08 &  3.10 & 0.00 \\

\teff\ + $\log g$ (A99, No Iter.) & 
0.203 $\pm$ 0.039 & 5.23 &  0.40 & 0.00 && 0.101 $\pm$ 0.045 & 2.25 &  6.66 & 5.55 \\

\teff\ + $\log g$ (A99, 5$^{\rm th}$ Iter.) & 
0.220 $\pm$ 0.043 & 4.19 &  0.90 & 0.00 && 0.129 $\pm$ 0.051 & 2.19 &  7.33 & 5.74 \\

Spectroscopic & 
0.194 $\pm$ 0.044 & 3.79 &  0.92 & 0.00 && 0.228 $\pm$ 0.053 & 3.74 &  1.09 & 0.00 \\

Fixed \teff &
0.191 $\pm$ 0.043 & 4.43 &  0.45 & 0.00 && 0.172 $\pm$ 0.060 & 2.86 &  3.42 & 2.86 \\

Fixed \vvhb & 
0.181 $\pm$ 0.034 & 5.36 &  0.41 & 0.00 && 0.132 $\pm$ 0.056 & 2.36 &  5.67 & 5.61 \\

\enddata
\tablenotetext{1}{Student's $t$-score.}
\tablenotetext{2}{The significance level to reject the hypothesis that
the mean \feh\ values of the \caw\ and \cas\ groups are identical.}
\tablenotetext{3}{The significance level to reject the hypothesis that
the mean \feh\ values of the \caw\ and \cas\ groups are identical from
bootstrap realization.}
\end{deluxetable}
\end{landscape}

\clearpage

\begin{landscape}
\begin{deluxetable}{cccccccccc}
\tablecaption{Difference in the mean iron abundances between M55 and NGC~6752.\label{tab:mock}}
\tabletypesize{\scriptsize}
\tablenum{3}
\tablewidth{0pc}
\tablehead{
\multicolumn{1}{c}{Method} &
\multicolumn{1}{c}{gf\tablenotemark{1}} &
\multicolumn{2}{c}{M55} &
\multicolumn{1}{c}{} &
\multicolumn{2}{c}{NGC~6752} &
\multicolumn{1}{c}{} &
\multicolumn{2}{c}{{\rm $\Delta$}\tablenotemark{2}} \\
\cline{3-4}\cline{6-7}\cline{9-10}
\multicolumn{1}{c}{} &
\multicolumn{1}{c}{} &
\multicolumn{1}{c}{\fehi} &
\multicolumn{1}{c}{\fehii} &
\multicolumn{1}{c}{} &
\multicolumn{1}{c}{\fehi} &
\multicolumn{1}{c}{\fehii} &
\multicolumn{1}{c}{} &
\multicolumn{1}{c}{\dfehi} &
\multicolumn{1}{c}{\dfehii} }
\startdata
Phot.\tablenotemark{3}  & 
C09 (No) &
 $-$2.087 $\pm$ 0.018 &  $-$1.936 $\pm$ 0.020 &&  $-$1.645 $\pm$ 0.014 &  $-$1.471 $\pm$ 0.015 && 
  0.443 $\pm$ 0.023 &   0.465 $\pm$ 0.025 \\ 
& C09 (5$^{\rm th}$) &
 $-$2.090 $\pm$ 0.016 &  $-$1.927 $\pm$ 0.028 &&  $-$1.642 $\pm$ 0.014 &  $-$1.432 $\pm$ 0.024 && 
  0.448 $\pm$ 0.021 &   0.496 $\pm$ 0.037 \\ 
 & Y13 (No) &
 $-$2.067 $\pm$ 0.016 &  $-$1.896 $\pm$ 0.021 &&  $-$1.687 $\pm$ 0.005 &  $-$1.463 $\pm$ 0.006 &&
  0.379 $\pm$ 0.017 &   0.443 $\pm$ 0.022 \\
 & Y13 (5$^{\rm th}$) &
 $-$2.073 $\pm$ 0.014 &  $-$1.869 $\pm$ 0.029 &&  $-$1.683 $\pm$ 0.005 &  $-$1.414 $\pm$ 0.010 &&
  0.390 $\pm$ 0.015 &   0.454 $\pm$ 0.031 \\
&&&&&&&&& \\

Phot.\tablenotemark{4} & 
C09 (No) &
 $-$2.114 $\pm$ 0.017 &  $-$1.789 $\pm$ 0.022 &&  $-$1.645 $\pm$ 0.014 &  $-$1.471 $\pm$ 0.015 && 
  0.469 $\pm$ 0.022 &   0.318 $\pm$ 0.027 \\ 
& C09 (5$^{\rm th}$)  &
 $-$2.110 $\pm$ 0.016 &  $-$1.883 $\pm$ 0.030 &&  $-$1.642 $\pm$ 0.014 &  $-$1.432 $\pm$ 0.024 && 
  0.469 $\pm$ 0.021 &   0.451 $\pm$ 0.038 \\ 
 & Y13 (No)  &
 $-$2.093 $\pm$ 0.016 &  $-$1.748 $\pm$ 0.022 &&  $-$1.687 $\pm$ 0.005 &  $-$1.463 $\pm$ 0.006 &&
  0.406 $\pm$ 0.017 &   0.285 $\pm$ 0.023 \\
 & Y13 (5$^{\rm th}$) &
 $-$2.092 $\pm$ 0.015 &  $-$1.825 $\pm$ 0.031 &&  $-$1.683 $\pm$ 0.005 &  $-$1.414 $\pm$ 0.010 &&
  0.409 $\pm$ 0.016 &   0.411 $\pm$ 0.032 \\
&&&&&&&&& \\

Spec.  & 
C09 &
 $-$2.003 $\pm$ 0.020 &  $-$2.002 $\pm$ 0.019 &&  $-$1.583 $\pm$ 0.027 &  $-$1.584 $\pm$ 0.025 &&
  0.420 $\pm$ 0.033 &   0.418 $\pm$ 0.031 \\
 & Y13  &
 $-$1.952 $\pm$ 0.014 &  $-$1.956 $\pm$ 0.017 &&  $-$1.657 $\pm$ 0.003 &  $-$1.635 $\pm$ 0.004 && 
  0.295 $\pm$ 0.014 &   0.321 $\pm$ 0.017 \\ 
&&&&&&&&& \\
  
Fixed \teff  & 
C09 &
 $-$2.001 $\pm$ 0.019 &  $-$2.026 $\pm$ 0.030 &&  $-$1.579 $\pm$ 0.025 &  $-$1.455 $\pm$ 0.025 &&
  0.422 $\pm$ 0.031 &   0.571 $\pm$ 0.039 \\
 & Y13  &
 $-$1.966 $\pm$ 0.015 &  $-$1.890 $\pm$ 0.028 &&  $-$1.655 $\pm$ 0.003 &  $-$1.463 $\pm$ 0.010 &&
  0.311 $\pm$ 0.015 &   0.426 $\pm$ 0.030 \\
&&&&&&&&& \\
  
Fixed \vvhb  & 
C09 &
 $-$1.999 $\pm$ 0.009 &  $-$2.006 $\pm$ 0.027 &&  $-$1.651 $\pm$ 0.011 &  $-$1.467 $\pm$ 0.021 &&
  0.348 $\pm$ 0.014 &   0.539 $\pm$ 0.034 \\ 
 & Y13  &
 $-$2.005 $\pm$ 0.008 &  $-$1.943 $\pm$ 0.029 &&  $-$1.690 $\pm$ 0.004 &  $-$1.465 $\pm$ 0.010 &&
  0.315 $\pm$ 0.009 &   0.479 $\pm$ 0.031 \\
\enddata
\tablenotetext{1}{ C09 = \citet{carretta09uves}; Y13 = \citet{yong13}.}
\tablenotetext{2}{$\Delta$ = {\rm [Fe/H]}$_{\rm N6752}$ $-$ {\rm [Fe/H]}$_{\rm M55}$ }
\tablenotetext{3}{Photometric method using two separate relations.}
\tablenotetext{4}{Photometric method using a single relation.}
\end{deluxetable}
\end{landscape}


\clearpage

\clearpage

\clearpage

\clearpage

\clearpage

\clearpage

\clearpage

\clearpage

\clearpage

\clearpage

\clearpage

\clearpage

\clearpage

\clearpage

\clearpage

\clearpage

\clearpage

\clearpage

\end{document}